\documentclass[12pt,epsf,amssymb]{article} \textheight =23 truecm
\def\lsim{\raise0.3ex\hbox{$<$\kern-0.75em\raise-1.1ex\hbox{$\sim$}}}
\def\gsim{\raise0.3ex\hbox{$>$\kern-0.75em\raise-1.1ex\hbox{$\sim$}}}
\def\noi{\noindent} \def\nn{\nonumber} \def\bea{\begin{eqnarray}}
\def\eea{\end{eqnarray}} \def\beq{\begin{equation}}
\def\eeq{\end{equation}} 
\def\beeq{\begin{eqnarray}} \def\eeeq{\end{eqnarray}} \def\R{ {\rm R
\kern -.31cm I \kern .15cm}} \def\C{ {\rm C \kern -.15cm \vrule
width.5pt \kern .12cm}} \def\Z{ {\rm Z \kern -.27cm \angle \kern
.02cm}} \def\N{ {\rm N \kern -.26cm \vrule width.4pt \kern .10cm}}
\def\1{{\rm 1\mskip-4.5mu l} }

\usepackage{graphics} \usepackage{graphicx} \usepackage{epsfig}

\begin{document} 
\begin{center} 
{\large \bf  The decays $\overline{\bf B} \to {\bf D}^{\bf **}\pi$}\par \vskip 3 truemm
{\large \bf and the Isgur-Wise functions $\tau_{\bf 1/2}{\bf (w)}$, $\tau_{\bf 3/2}{\bf (w)}$} 

\vskip 1 truecm {\bf F. Jugeau, A. Le Yaouanc, L. Oliver and J.-C.
Raynal}\\

{\it Laboratoire de Physique Th\'eorique}\footnote{Unit\'e Mixte de
Recherche UMR 8627 - CNRS }\\    {\it Universit\'e de Paris XI,
B\^atiment 210, 91405 Orsay Cedex, France} \end{center}

\begin{abstract}  
We perform a phenomenological analysis of the decays $B \to D^{**}\pi$,
where $D^{**}$ is a $P$-wave excited meson with total angular momentum
$j = {1 \over 2}$ or ${3 \over 2}$ for the light cloud, recently
measured by the Belle Collaboration in the modes $\overline{B}^0\to
D^{**+}\pi^-$ (Class I) and $B^- \to D^{**0}\pi^-$ (Class III). Making
the reasonable assumption of naive factorization, that we test in $B \to D(D^*)\pi$ decays, Class~I decays allow to extract the
Isgur-Wise form factors $\tau_{1/2}(w)$, $\tau_{3/2}(w)$ at $w \cong
w_{max}$ ($q^2 \cong 0$).  We obtain $\tau_{1/2}(w_{max}) < 0.20$, $\tau_{3/2}(w_{max}) = 0.31 \pm 0.12$. We discuss the question of
the $w$ dependence of these IW functions. We find agreement with the
Bakamjian-Thomas quark model of form factors and, extrapolating at
$w=1$, with Bjorken and Uraltsev sum rules. We discuss also Class III
decays, where the $D^{**0}$ $(j = {1 \over 2})$ emission diagram contributes.  We extract the corresponding $f_{D_{1/2}}$ decay constant,
that is in agreement with theoretical estimates at finite mass. Finally, we must
warn that $1/m_Q$ corrections could be large and upset the results of
the present stage of this analysis. On the other hand, we confront
present data on the semileptonic rate of $B$ mesons to excited states with theoretical expectations.  \end{abstract}

\vskip 1 truecm

\noi LPT Orsay 05-07 \par 
\noi February 2005\par \vskip 1 truecm

\noindent e-mails : frederic.jugeau@th.u-psud.fr,
leyaouan@th.u-psud.fr, oliver@th.u-psud.fr 

\newpage \pagestyle{plain}
\section{Introduction.} \hspace*{\parindent} 
Our main purpose in this paper is to extract information on the lowest
$(n=0)$ heavy quark Isgur-Wise functions $\tau_{1/2}(1)$ and
$\tau_{3/2}(1)$, that correspond to the transitions ${1\over 2}^- \to
{1 \over 2}^+$ or ${3 \over 2}^+$, quantities that are of importance in
heavy quark physics, for instance in small velocity sum rules (SR). We will first make a naive estimation of these
quantities from $\overline{B}^0 \to D^{**+}\pi^-$ decays assuming factorization and the
heavy quark limit. We will moreover discuss the more
involved decays $B^- \to D^{**0}\pi^-$ in which there is $D^{**0}$ emission, in
particular the sign of the interference between the $\pi^-$ and $D^{**0}$
emission diagrams. \par

There are four $P$-wave $D^{**}$ mesons corresponding to the coupling
of the total quark spin $S = 0, 1$ and the orbital momentum $\ell = 1$. In
the language of the heavy quark limit, where the total angular momentum
$j$ of the light quark is a good quantum number ($j = {1 \over 2}$ or
${3 \over 2}$), these states can be denoted by 
$D_J^j$ where $J$ is the total angular momentum of the state. There are
then four possibilities $(j,J) = \left ( {1 \over 2}, 0\right )$,
$\left ( {1 \over 2}, 1\right )$, $\left ( {3 \over 2}, 1\right )$ or $\left
({3 \over 2}, 2\right )$.\par

According to the states of charge, the $B \to D^{**}\pi$ decays are of three classes,
following the classification of B. Stech and collaborators
\cite{1r}, \cite{2r}~: \par

(i) $\overline{B}^0 \to D^{**+}\pi^-$ (Class I) where only the $\pi$
emission diagram contributes~;\par

(ii) $\overline{B}^0 \to D^{**0}\pi^0$ (Class II) where only the
$D^{**0}$ diagram contributes~;\par

(iii) $B^- \to D^{**0}\pi^-$ (Class III), where both diagrams
of $\pi$ emission and $D^{**0}$ emission contribute.\par

It is worth to recall that isospin symmetry relates the amplitudes of
the three classes in this particular case, namely \cite{3r}~:

\beq \label{1e} A\left ( B^- \to D^{**0} \pi^-\right ) = A \left (
\overline{B}^0 \to D^{**+}\pi^-\right ) - \sqrt{2} A \left (
\overline{B}^0 \to D^{**0}\pi^0\right ) \ . 
\eeq

The Belle Collaboration has obtained in the past very interesting
results on Class~III decays $B^- \to D^{**0}\pi^-$ \cite{4r}. Four
states where indeed observed, two narrow states corresponding to the $j
= {3 \over 2}$ states, that decay into $D(D^*)$ in the $D$-wave, and
two very wide states that decay into $D(D^*)$ in the $S$-wave. \par

The interesting news is that recently, at the Beijing ICHEP
04 Conference, the Belle Collaboration has presented results on 
Class I decays $\overline{B}^0 \to D^{**+}\pi^-$ \cite{5r}. Interestingly, 
the narrow states have been observed, while only limits on the decays into wide
states have been obtained, indicating a much smaller BR. The Belle data have recently drawed the attention of the theory \cite{reference6}.\par

An enormous theoretical effort has been dedicated in the last five
years to the understanding of non-leptonic two-body $B$ decays in the
cases of the {\it emission} of a light meson like $\pi$ or $\rho$ in
the so-called QCD Factorization approach \cite{6newref}, in the
Perturbative QCD Factorization approach
\cite{7newref} or within the Soft Collinear Effective Theory \cite{reference8}. These methods have been applied to two-body decays into
ground state mesons. In the present paper we are dealing with decays
into excited $D^{**}$ mesons, with both $\pi$ or $D^{**}$ emission
diagrams. For Class I decays we have only the $\pi$ emission diagram,
and in this case we could in principle use the QCD methods of these
papers. However, in Class III decays there is the diagram of $D^{**}$
emission, a meson composed of heavy-light quarks, for which there are no rigorous results. Moreover, we are
dealing with the first measurements of these decays, that hopefully will be
refined in the future. For
these reasons, as a preliminary study, we will stick to the naive
factorization approach \cite{1r}, \cite{2r} in order to investigate if
there is a sensible description of the decays $B \to D^{**}\pi$ within this simple phenomenological approach. \par

Class I decays and the $\pi$ emission diagram of Class III 
are related to the $B \to D^{**}$ form factors that, in the heavy quark
limit, reduce to two Isgur-Wise (IW) functions $\tau_{1/2}(w)$ and
$\tau_{3/2} (w)$ \cite{6r}. These form factors are of a significant
theoretical importance, since they are related, at zero recoil $w=1$,
to the slope of the elastic IW function through Bjorken \cite{6r},
\cite{7r} and Uraltsev SR \cite{8r}. \par

In a recent paper we have tried to use the Belle data on Class III
decays to extract $\tau_{1/2}(w_{max})$ and $\tau_{3/2}(w_{max})$,
where $w(q^2 \cong 0) \cong w_{max}$ \cite{9r}. This calculation relied
on a strong hypothesis, namely that the diagram of $D^{**0}$ emission should be
small. We got some results on $\tau_{1/2}(w_{max})$,
$\tau_{3/2}(w_{max})$ that were extrapolated to $\tau_{1/2}(1)$,
$\tau_{3/2}(1)$ assuming the $w$-dependence of the form factors given
by the Bakamjian-Thomas (BT) class of relativistic quark models that
yield covariant form factors exhibiting heavy quark symmetry
\cite{10r}. We obtained
$\tau_{1/2}(1) \sim \tau_{3/2}(1)$, at odds with the expectations of
Uraltsev SR.\par

However, to neglect the $D^{**0}$ diagram is a rough
approximation that could be unfounded \cite{11r}. It is well-known that
in some cases the color-suppressed diagrams like the $D^{**0}$
emission one are often not as suppressed as one could expect on naive
grounds. Therefore, our determination of $\tau_{1/2}(w_{max})$,
$\tau_{3/2}(w_{max})$ has to be reconsidered using only Class I decays,
now measured, and where only the $\pi$ emission diagram
contributes. Hence the interest of the new results on Class I
decays on which we will first concentrate. Below we will come
back to the interpretation of the results on Class III decays,
taking into account $D^{**0}$ emission. In what follows there is some unavoidable overlap with our Appendix B
of ref. \cite{9r} from which some points are worth to be recalled.\par

The paper is organized as follows. In Section 2 we extract
$\tau_{1/2}(w_{max})$, $\tau_{3/2}(w_{max})$ from Class I decays
$\overline{B}^0 \to D^{**+}\pi^-$ using factorization, and discuss the
question of the $w$-dependence and the extrapolation of
$\tau_{1/2}(w)$, $\tau_{3/2}(w)$ at $w=1$, the comparison with Bjorken and Uraltsev SR and with BT quark
models. In Section 3
we combine the results of Section 2 with the measured rates of Class III decays $B^- \to
D^{**0}\pi^-$ in order to extract the needed value for the decay constant
$f_{D_{1/2}}$, non-vanishing in the heavy quark limit, and we
compare with theoretical predictions. We treat with
special care the question of the interference between the $\pi^-$ and $D^{**0}$ emission diagrams. In Section 4
we make predictions for Class II decays $\overline{B}^0 \to
D^{**0}\pi^0$. In Section 5 we discuss the implications for
the semileptonic decays $\overline{B}^0 \to D^{**+}\ell^-\overline{\nu}$ and compare with the scarce existing data, and in Section 6 we conclude. In Appendix A we reproduce the Belle data for Class I
and Class III decays to make clear in the text how we extract the
rates $B \to D^{**}\pi$ of the different modes and how we treat the
experimental errors. In Appendix B we use the factorization model to
compare with the data on $B \to D(D^*)\pi$, where all modes have been
measured, and we extract the effective coefficients $a_1$ and $a_2$
that enter in the factorization model. In Appendix C we discuss the
corrections to factorization and finally Appendix D is devoted to the
question of the $1/m_Q$ corrections to the heavy quark limit.

\section{Extraction of $\tau_{\bf 1/2}({\bf w}_{\bf max})$, $ \tau_{\bf
3/2}({\bf w}_{\bf max})$ from Class I decays.} \hspace*{\parindent}
Let us now consider the Class I decays measured by the Belle Collaboration,
where for the wide states the masses of the results of Class III decays
are assumed (Table 5 of Appendix A). \par

Assuming that these states decay essentially into two-body modes, i.e.\break\noindent
$B\left ( D_2^{3/2} \to (D + D^*)\pi \right )$, $B\left ( D_1^{3/2} \to
D^*\pi \right )$, $B\left ( D_0^{1/2} \to D\pi \right )$, $B\left (
D_1^{1/2} \to D^*\pi \right )$, the following branching ratios are
given by a Clebsch-Gordan coefficient
\beq \label{2e} B\left ( D_1^{3/2\ +} \to D^{*0}\pi^+ \right )= B\left
( D_0^{1/2\ +} \to D^0\pi^+ \right ) = B\left ( D_1^{1/2\ +} \to
D^{*0}\pi^+ \right ) = {2 \over 3}\ . 
\eeq

To estimate $B\left ( D_2^{3/2\ +} \to D^0\pi^+ \right )$ and $B\left (
D_2^{3/2\ +} \to D^{*0}\pi^+ \right )$, we use the spin counting of the
non-relativistic quark model. {\it In the limit of heavy quark symmetry}, i.e. assuming the pairs $(D
,D^*)$, $(D_2^{3/2}, D_1^{3/2})$ and $(D_1^{1/2}, D_0^{1/2})$ to be degenerate, simple angular
momentum calculations give
\beq \label{3e} \Gamma \left ( D_2^{3/2}\right ) = \Gamma \left (
D_1^{3/2}\right ) \qquad \qquad \Gamma \left ( D_0^{1/2}\right ) = \Gamma
\left ( D_1^{1/2}\right ) \eeq

\noi
\beq \label{4e} \Gamma \left ( D_2^{3/2} \to D^* \pi \right ) = {3
\over 2} \ \Gamma \left ( D_2^{3/2}\to D\pi \right ) \ . \eeq

\noi This last relation gives the needed spin counting coefficient.
\par

It is easy to obtain this factor by realizing that to have the $D$ wave
one needs (1 denoting the quark emitting a pion) the operator (taking
Oz along the pion momentum) to emit a pion in the $D$ wave reads
$(\sigma_1^z k_{\pi}) \exp (iz_1k_{\pi}) \to i k_{\pi}^2 \sigma_1^z
z_1$. We have then, for the non-vanishing amplitudes 
$$M \left ( D_2^{3/2} \to D\pi \right ) = \ <1\ 0, 1\ 0|2\ 0>\ <0\ 0|\sigma_1^z|1\ 0>\ <0\ 0|{\cal Y}_1^z |1\ 0>$$
\beq
\label{5e}
M \left ( D_2^{3/2(\pm 1)} \to D^{*(\pm 1)}\pi \right ) = \ <1\ 0, 1\pm 1|2\pm 1>\ <1\pm 1|\sigma_1^z|1\pm 1>\ <0\ 0|{\cal Y}_1^z |1\ 0>
\eeq

\noi that gives
\bea
\label{6e}
&&M \left ( D_2^{3/2} \to D\pi \right ) = \sqrt{{2 \over 3}} \ <0\ 0 |{\cal Y}_1^z|1\ 0>\nn \\
&&M \left ( D_2^{3/2(\pm 1)} \to D^{*(\pm 1)}\pi \right ) = \pm {1 \over \sqrt{2}} \ <0\ 0 |{\cal Y}_1^z|1\ 0>
\eea

\noi and hence (\ref{4e}).\par

We now take into account the actual masses. Since both $D_2^{3/2} \to
D\pi$ and $D_2^{3/2} \to D^*\pi$ proceed through the $D$-wave, we will
have, in an obvious notation, in the isospin symmetry limit,
\beq
\label{7e}
{\Gamma \left ( D_2^{3/2\ +} \to D^0 \pi^+ \right ) \over \Gamma \left (
D_2^{3/2\ +} \to D^{*0} \pi^+ \right )} = {\Gamma \left ( D_2^{3/2\ 0} \to
D^+ \pi^- \right ) \over \Gamma \left ( D_2^{3/2\ 0} \to D^{*+} \pi^-
\right )} = {2 \over 3} \ {p^5 \over p^{*5}} \cong 2.5 \ .
\eeq

\noi This estimation is in agreement with the present world averages \cite{12r}
\bea
\label{8e}
&&{\Gamma \left ( D_2^{3/2\ +} \to D^0 \pi^+ \right ) \over \Gamma \left (
D_2^{3/2\ +} \to D^{*0} \pi^+ \right )} = 1.9 \pm 1.1 \pm 0.3\nn \\
&&{\Gamma \left ( D_2^{3/2\ 0} \to
D^+ \pi^- \right ) \over \Gamma \left ( D_2^{3/2\ 0} \to D^{*+} \pi^-
\right )} = 2.3 \pm 0.6 \ .
\eea

\noi Therefore, we obtain the branching ratios
\beq
\label{9e}
B \left ( D_2^{3/2\ +} \to D^0 \pi^+ \right ) \cong 0.48 \qquad \qquad B\left (
D_2^{3/2\ +} \to D^{*0} \pi^+ \right ) \cong 0.19\ . 
\eeq

\noi From these BR, adding the errors in quadrature, we find roughly
\bea \label{10e} 
&B\left ( \overline{B}^0 \to D_2^{3/2\ +}\pi^-\right ) =
(6.4 \pm 0.8) \times 10^{-4} &\qquad \left ( {\rm from}\  D_2^{3/2\ +} \to D^0 \pi^+\right )
\nn \\ 
&B\left ( \overline{B}^0 \to D_2^{3/2\ +}\pi^-\right ) = (12.9 \pm
3.3) \times 10^{-4} &\qquad \left ( {\rm from}\  D_2^{3/2\ +} \to D^{*0} \pi^+\right )   \eea

We realize that the $B\left ( \overline{B}^0 \to D_2^{3/2\ +}\pi^-\right
)$ differs if one obtains it from $D_2^{3/2\ +} \to D^0\pi^+$ or from
$D_2^{3/2\ +} \to D^{*0}\pi^+$, and the values are consistent only within $2 \sigma$.
Using (\ref{2e}) for the other modes, and taking into account the large
uncertainty from both results (\ref{10e}) one finds 
\bea
\label{11e}
&&B\left ( \overline{B}^0 \to D_2^{3/2\ +}\pi^-\right ) = (10.9 \pm 5.3) \times 10^{-4} \nn \\
&&B\left ( \overline{B}^0 \to D_1^{3/2\ +}\pi^-\right ) = (5.5 \pm 1.4) \times 10^{-4} \nn \\
&&B\left ( \overline{B}^0 \to D_0^{1/2\ +}\pi^-\right ) < 1.8 \times 10^{-4} \nn \\
&&B\left ( \overline{B}^0 \to D_1^{1/2\ +}\pi^-\right ) < 1.0 \times 10^{-4}\ . 
\eea

Assuming factorization of $\pi^-$ emission, as it is reasonable within
the BBNS QCD factorization scheme in the heavy quark limit
\cite{6newref}, and assuming that the states $1^+$ are unmixed, we find for
the decay rates
\beq
\label{12e}
\Gamma = {G_F^2 \over 16 \pi} \left | V_{cb} V_{ud}^*\right |^2 {p \over m_B^2} \left | M(B \to D^{**}\pi ) \right |^2 
\eeq

\noi where
\bea
\label{13e}
&&\left | M\left ( \overline{B}^0 \to D_2^{3/2\ +} \pi^- \right ) \right |^2 = 2m_Dm_B
\left ( m_B + m_D\right )^2 \left ( w_0^2 - 1 \right )^2 a_1^2 f_{\pi}^2
\left | \tau_{3/2}(w_0)\right |^2 \nn \\
&&\left | M\left ( \overline{B}^0 \to D_1^{3/2\ +} \pi^- \right ) \right |^2 = 2m_Dm_B
\left ( m_B - m_D\right )^2 \left ( w_0 + 1 \right )^2 \left ( w_0^2 - 1 \right )a_1^2 f_{\pi}^2
\left | \tau_{3/2}(w_0)\right |^2 \nn \\
&&\left | M\left ( \overline{B}^0 \to D_1^{1/2\ +} \pi^- \right ) \right |^2 = 4m_Dm_B
\left ( m_B - m_D\right )^2 \left ( w_0^2 - 1 \right ) a_1^2 f_{\pi}^2
\left | \tau_{1/2}(w_0)\right |^2 \nn \\
&&\left | M\left ( \overline{B}^0 \to D_0^{1/2\ +} \pi^- \right ) \right |^2 = 4m_Dm_B
\left ( m_B + m_D\right )^2 \left ( w_0 - 1 \right )^2 a_1^2 f_{\pi}^2
\left | \tau_{1/2}(w_0)\right |^2 
\eea

\noi with
\beq
\label{14e}
w_0 \cong {m_B^2 + m_{D}^2 \over 2m_B m_{D}}
\eeq

\noi the subindex 0 denoting the value of $w$ for $q^2 = m_{\pi}^2
\cong 0$ and $m_{D}$ the mass of the corresponding $D_J^j$ state. The short distance QCD factor $a_1$ is close to 1 (Appendix B).\par

It is interesting to notice that the rates (\ref{12e}), (\ref{13e}) are
given, assuming the $D^{**}$ for a given $j = {1 \over 2}$ or ${3 \over
2}$ to be degenerate, by the expressions 
\bea
\label{15e}
&&\Gamma \left ( \overline{B}^0 \to D_2^{3/2\ +} \pi^- \right )  = \Gamma \left ( \overline{B}^0 \to D_1^{3/2\ +} \pi^- \right )\nn \\
&&= {G_F^2 \over 16 \pi} \left | V_{cb} V_{ud}^*\right |^2 m_B^3\ a_1^2\ f_{\pi}^2 {(1 - r)^5 (1+r)^7 \over 16r^3} \left | \tau_{3/2} \left ( {1 + r^2 \over 2r}\right ) \right |^2
\eea
\bea
\label{16e}
&&\Gamma \left ( \overline{B}^0 \to D_1^{1/2\ +} \pi^- \right )  = \Gamma \left ( B \to D_0^{1/2} \pi \right )\nn \\
&&= {G_F^2 \over 16 \pi} \left | V_{cb} V_{ud}^*\right |^2 m_B^3\ a_1^2\ f_{\pi}^2 {(1 - r)^5 (1+r)^3 \over 2r} \left | \tau_{1/2} \left ( {1 + r^2 \over 2r}\right ) \right |^2
\eea

\noi where $r = {m_D^{3/2} \over m_B}$ and $r = {m_D^{1/2} \over m_B}$
respectively in the first and second relations. The equalities $\Gamma \left ( B \to D_2^{3/2} \pi \right )  = \Gamma
\left ( B \to D_1^{3/2} \pi \right )$, $\Gamma \left ( B \to
D_1^{1/2} \pi \right )  = \Gamma \left ( B \to D_0^{1/2} \pi \right )$
follow from heavy quark symmetry since the $B$ meson is spinless,
and there is a single helicity amplitude for each decay, 
the emission of a longitudinally polarized $D^{**}$. \par

Using the central values for the masses, but taking into account the
errors in (\ref{11e}) and $|V_{cb}| = 0.040 \pm 0.002$ (Appendix B), we
find from the different modes,
\bea
\label{17e}
&B \to D_2^{3/2} \pi &\qquad \qquad \left | \tau_{3/2}(1.31)\right | = 0.32 \pm 0.10 \nn \\
&B \to D_1^{3/2} \pi &\qquad \qquad \left | \tau_{3/2}(1.32)\right | = 0.23 \pm 0.04 \nn \\
&B \to D_0^{1/2} \pi &\qquad \qquad \left | \tau_{1/2}(1.37)\right | < 0.20 \nn \\
&B \to D_1^{1/2} \pi &\qquad \qquad \left | \tau_{1/2}(1.32)\right | < 0.16 \ .
\eea

Within $1 \sigma$ there is consistency between the different
determinations of $|\tau_{3/2}(w_0)|$ and $|\tau_{1/2}(w_0)|$, but
errors increase considering both determinations. Since besides the
statistical errors, there are systematic errors (from experiment and
theory), we consider safer to take the union of the domains (\ref{17e})
rather than their intersection. We conclude safely that we will have
the numbers
\bea
\label{18e}
&&\left | \tau_{3/2}(1.31)\right | = 0.31 \pm 0.12 \nn \\
&&\left | \tau_{1/2}(1.37)\right | < 0.20 \ .
\eea

\subsection{Extrapolation at w = 1 and comparison with
Bjorken and Uraltsev sum rules.} \hspace*{\parindent}
Our results have been obtained at $w_{max}$. It is of interest to know
the values of $\tau_{1/2}(w)$, $\tau_{3/2}(w)$ at $w = 1$. There are a
number of values for the slopes of these IW functions in the
literature. \par

We should keep in mind that we have two rather loose constraints on
$\tau_{1/2}(w)$, $\tau_{3/2}(w)$, namely the values at $w_{max}$
(\ref{18e}) and the qualitative idea that the $n=0$ IW functions should
give a main contribution to Bjorken and Uraltsev 
sum rules.\par

Let us consider, {\it as an illustration}, the
parametrization obtained within BT quark models (last reference \cite{10r})
\bea
\label{19e}
&\tau_{1/2}(w) = \tau_{1/2}(1) \left ( {2 \over w+1}\right )^{2\sigma_{1/2}^2} &\qquad\qquad \sigma_{1/2}^2 = 0.83 \nn \\
&\tau_{3/2}(w) = \tau_{3/2}(1) \left ( {2 \over w+1}\right )^{2\sigma_{3/2}^2} &\qquad\qquad \sigma_{3/2}^2 = 1.5
\eea

\noi we obtain at zero recoil
\bea
\label{20e}
&&\left | \tau_{3/2}(1) \right | = 0.46 \pm 0.18 \nn \\
&&\left | \tau_{1/2}(1) \right | <  0.26
\eea

\noi to be compared with the values in the BT model
\bea
\label{21e}
&&\left | \tau_{3/2}(1) \right |^{BT} = 0.54 \nn \\
&&\left | \tau_{1/2}(1) \right |^{BT} =  0.22\ . 
\eea

\noi We find agreement for $|\tau_{3/2}(1)|$ within errors, and
$|\tau_{1/2}(1)|$ could still be roughly consistent with the BT
model.\par

Bjorken \cite{6r}, \cite{7r} and Uraltsev \cite{8r} sum rules write, respectively
\begin{eqnarray}
\label{equationnumber22}
&&\rho^2 = {1 \over 4} + \sum_n \left | \tau_{1/2}^{(n)}(1)\right |^2 + 2 \sum_n \left | \tau_{3/2}^{(n)}\right |^2 \nn \\
&&\sum_n \left | \tau_{3/2}^{(n)}(1) \right |^2 - \sum_n \left | \tau_{1/2}^{(n)} (1) \right |^2 = {1 \over 4}\ .
\end{eqnarray}

It is understood that these SR are truncated at some $n$ that
corresponds to a scale $\Delta \sim$ 1~GeV and it is then natural to
assume that the ground state dominates. Keeping thus the $n = 0$
states, with which we are dealing here, we get contributions to Bjorken
and Uraltsev SR that lie in the following ranges~:
\bea
\label{22e}
&&0.40 < {1 \over 4} + \left | \tau_{1/2}(1) \right |^2 + 2 \left | \tau_{3/2}(1)\right |^2 < 1.11 \nn \\
&&0.03 < \left | \tau_{3/2}(1) \right |^2 - \left | \tau_{1/2}(1)\right |^2 < 0.41 \ .
\eea

\noi Therefore, the $n = 0$ states could give an important contribution to
the SR and, considering this piece as dominant, low values for
$\rho^2$ are not excluded nor the value ${1 \over 4}$ for the r.h.s. of
Uraltsev SR.\par

There are other theoretical estimates of the IW functions
$\tau_{1/2}(w)$, $\tau_{3/2}(w)$, mainly within the QCD Sum Rules
approach \cite{23r}, \cite{24r}, \cite{25newref}, \cite{26newref}. The
pioneering calculations of $\tau_{1/2}(w)$ and $\tau_{3/2}(w)$
\cite{24r} show indeed that at large $w$ the slope of $\tau_{1/2}(w)$
is much smaller than the one of $\tau_{3/2}(w)$, as in the BT model,
while the values at $w=1$ were found roughly equal for both IW functions,
$\tau_{1/2}(1) = \tau_{3/2}(1) \cong 0.24$. On the other hand,
next-to-leading calculations of the function $\tau_{1/2}(w)$  have
later been performed \cite{25newref}, giving $\tau_{1/2}(1) = 0.35 \pm
0.10$ and a slope of the order $\sigma_{1/2}^2 = 0.5$, in our notation
(\ref{19e}). In view of the importance of the corrections, it would be
interesting to have the corresponding calculation for $\tau_{3/2}$.
This latter value for $\tau_{1/2}(1)$ is larger than the value obtained
in the present paper. On the other hand, in ref. \cite{26newref} there
is a calculation of both $\tau_{1/2}(w)$ and $\tau_{3/2}(w)$ ($\zeta
(w) = 2 \tau_{1/2}(w)$, $\tau (w) = \sqrt{3}\tau_{3/2}(w)$ \cite{32r})
and gives $\tau_{3/2}(1) = 0.43 \pm 0.08$, $\tau_{1/2}(1) = 0.13 \pm
0.04$ and the slopes $\sigma_{3/2}^2 = 0.90 \pm 0.05$, $\sigma_{1/2}^2
= 0.50 \pm 0.05$. These results imply a sizeable contribution to
Uraltsev SR, are in agreement with the determinations of the present
paper, and are qualitatively consistent with the BT model results
(\ref{21e}) and with (\ref{19e}), $\tau_{3/2}(w)$ being steeper than
$\tau_{1/2}(w)$. One should notice that a different interpolating field
for $\tau_{1/2}$ is used by \cite{26newref} from the one in \cite{24r},
\cite{25newref}, and that radiative corrections are absent. Recently, a
lattice determination has obtained the values $\tau_{1/2}(1) = 0.38(5)$
and $\tau_{3/2}(1) = 0.53(8)$, with unknown systematic errors
\cite{29newref}. These values imply a sizeable contribution to Uraltsev
SR. Compared with the BT determination (\ref{21e}), $\tau_{3/2}(1)$ is
in fair agreement, while $\tau_{1/2}(1)$ is larger, and in agreement
with the QCDSR determination \cite{25newref}. A fortiori, this latter
value is much larger than the QCDSR result \cite{26newref}, and also
than the present phenomenological limit (\ref{18e}) obtained in the
present paper, with the extrapolation from $w_{max}$ assumed here. We summarize the situation in Table 1.
\vskip 5 truemm
\begin{center}
\begin{tabular}{|c|c|c|c|c|}
\hline
Theoretical  &$\tau_{1/2}(1)$ &$\sigma_{1/2}^2$ &$\tau_{3/2}(1)$ &$\sigma_{3/2}^2$\\
method & & & & \\
\hline
QCDSR &$0.35 \pm 0.10$ &$0.5$ & & \\
(NLO) \cite{25newref} & & & & \\
\hline
QCDSR \cite{26newref} &$0.13 \pm 0.04$ &$0.50 \pm 0.05$ &$0.43 \pm 0.08$ &$0.90 \pm 0.05$\\
\hline
BT Quark &$0.22$ &$0.83$ &$0.54$ &$1.5$ \\
Model \cite{10r} & & & &\\
\hline
Lattice \cite{29newref} &$0.38(5)$ & &$0.53(8)$ &\\
 \hline
Present &$< 0.26$ &$0.83$ &$0.46 \pm 0.18$ &$1.5$\\
paper & &(input) & &(input)\\
 \hline
\end{tabular}
\end{center}
\noi {\bf Table 1.} The values at zero recoil $\tau_j(1)$ and slopes $\sigma_j^2$ $(j =
{1 \over 2}, {3\over 2})$ in the different theoretical approaches,
compared with the phenomenological determination at $w_{max}$ of present paper, extrapolated at $w=1$ with the slopes of the BT quark model (\ref{19e}).\par
\vskip 5 truemm

\section{Interference with ${\bf D}^{\bf **0}$ emission in Class
III\break\noindent decays.} \hspace*{\parindent}
Let us now consider Class III decays measured by the Belle Collaboration,
that we summarize in Table 6 of Appendix A. \par

We have here the same BR as for the modes of Class I, namely
\bea
\label{24e}
&&B\left ( D_1^{3/2\ 0} \to D^{*+}\pi^-\right ) = B \left ( D_0^{1/2\ 0} \to D^{+}\pi^-\right ) = B\left ( D_1^{1/2\ 0} \to D^{*+}\pi^-\right ) = {2 \over 3}\nn \\
&&B\left ( D_2^{3/2\ 0} \to D^{+}\pi^-\right ) \cong 0.48 \qquad \qquad B\left ( D_2^{3/2\ 0} \to D^{*+}\pi^-\right ) \cong 0.19\ .
\eea

We find, adding the errors in quadrature
\bea
\label{25e}
&B \left ( B^- \to D_2^{3/2\ 0} \pi^-\right ) = (7.1 \pm 1.6) \times 10^{-4} &\qquad \left ( {\rm from}\ D_2^{3/2\ 0} \to D^+\pi^-\right ) \nn \\
&B \left ( B^- \to D_2^{3/2\ 0} \pi^-\right ) = (9.5 \pm 2.5) \times 10^{-4} &\qquad \left ( {\rm from}\ D_2^{3/2\ 0} \to D^{*+}\pi^-\right ) .
\eea

We realize that the values obtained for $B(B^- \to D_2^{3/2\ 0} \pi^-)$
from $D_2^{3/2\ 0} \to D^+\pi^-$ or $D_2^{3/2\ 0} \to D^{*+}\pi^-$
agree within $1\sigma$. Using (\ref{24e}) for the other modes, and
taking into account the uncertainty from both results (\ref{25e}) one
finds
\bea
\label{26e}
&&B \left ( B^- \to D_2^{3/2\ 0} \pi^-\right ) = (8.7 \pm 3.2) \times 10^{-4} \nn \\
&&B \left ( B^- \to D_1^{3/2\ 0} \pi^-\right ) = (10.2 \pm 2.3) \times 10^{-4}\nn \\
&&B \left ( B^- \to D_0^{1/2\ 0} \pi^-\right ) = (9.1 \pm 2.9)\times 10^{-4}\nn \\
&&B \left ( B^- \to D_1^{1/2\ 0} \pi^-\right ) = (7.5 \pm 1.7) \times 10^{-4} \ .  
\eea

Comparing these BR with the corresponding Class I (\ref{11e}) we
see that there is a large difference for the $j = {1 \over 2}$
states, while there is consistency for the $j = {3 \over 2}$ states.
This is interesting and seems to indicate that the $D^{**0}$ emission
diagram could be very important \cite{11r}. This is likely,
because the decay constants of $j =
{1 \over 2}$ states do not vanish, while those of $j = {3\over 2}$
states vanish in the heavy quark limit \cite{18r} \cite{19newreference} \cite{3r}
\beq
\label{27e}
f_{1/2} \not= 0 \qquad\qquad \qquad \qquad f_{3/2} = 0\ .
\eeq

As demonstrated in \cite{18r}, the equality $f_{3/2} = 0$ follows
intuitively from the fact that the multiplet $j = {3 \over 2}$ contains
two states with $J = 1,2$ and there is no current coupling the vacuum to $J = 2$.
On the other hand, $f_{1/2} \not= 0$ follows because the $D^{**}(J=0)$
is a system of widely unequal masses, and vector current conservation
does not hold. We assume, following \cite{11r}, that the decays $B^- \to
D_0^{1/2\ 0} \pi^-$ and $B^- \to D_1^{1/2\ 0} \pi^-$ have a sizeable
contribution from $D_J^{1/2\ 0}$ ($J = 0,1)$ emission via,
respectively, the vector and axial current. \par

We now consider both diagrams for Class III decays and we will
take care of the delicate question of the relative sign between the
$\pi$ emission and the $D^{**0}$ emission diagrams.

The decays $B \to D_0^{1/2}\pi$ and $B \to D_1^{1/2}\pi$ are
respectively $S$-wave parity conserving and $P$-wave parity violating.
Let us define in an homogeneous way the needed matrix elements $(q = p
- p')$ \cite{18r}. For $\pi$ emission we need the current matrix elements

$$<\pi (q)|A_{\mu}|0> \ = f_{\pi} \ q_{\mu}$$
$$<D_0^{1/2\ 0}(p') |A_{\mu}|B^-(p)>\ = \sqrt{m_Dm_B}\ 2(v'-v)_{\mu} \ \tau_{1/2}(w)$$
\beq
\label{28e}
<D_1^{1/2\ 0}(p', \varepsilon ) |V_{\mu}|B^-(p)>\ = \sqrt{m_Dm_B}\ 2\left [ (w-1) \varepsilon_{\mu}^* - (\varepsilon^* \cdot v)v'_{\mu}\right ] \tau_{1/2}(w) 
\eeq

\vskip 5 truemm
\noi while for $D^{**}$ emission \cite{18r}, \cite{14r}
$$<D_0^{1/2\ 0}(p') |V_{\mu}|0>\ = f_{D_{1/2}} \ p'_{\mu}$$
$$<D_1^{1/2\ 0}(p', \varepsilon ) |A_{\mu}|0>\ = - f_{D_{1/2}}\ m_{D_{1/2}}\ \varepsilon_{\mu}^*$$
\beq
\label{29e}
<\pi^-(q)|V_{\mu}| B^-(p)>\ = \left ( p_{\mu} + q_{\mu} - {m_B^2 - m_{\pi}^2 \over p'^2} p'_{\mu}\right ) f_+^{\pi B}(p'^2) + {m_B^2 - m_{\pi}^2 \over p'^2} p'_{\mu} f_0^{\pi B} (p'^2)\ .
\eeq

The minus sign for the definition of the $D_1^{1/2}$ decay constant
comes from the Clebsch-Gordan convention of
Isgur and Wise (coupling the orbital angular momentum $\ell = 1$ with
the light quark spin $s_q = {1 \over 2}$ to give $j = {1 \over 2}$)
that yields the definitions (\ref{28e}). From
(\ref{29e}), as predicted by heavy quark symmetry, one obtains,
\beq
\label{30e}
<D_0^{1/2\ 0}(p') |A^3|B^-(p)>\ = - <D_1^{1/2\ 0}(p', \varepsilon ) |V^0|B^-(p)>\ = \sqrt{m_Dm_B}\ 2v'^3 \tau_{1/2}(w)
\eeq

\noi and corresponds to the convention
\beq
\label{31e}
S_3^c |D_1^{{1\over 2}^+}>\ = - |D_0^{{1 \over 2}^+}>\ .
\eeq

\noi Likewise, one must have
\beq
\label{32e}
<D_0^{1/2\ 0}(p') |V^3|0>\ = - \ <D_1^{1/2\ 0}(p', \varepsilon ) |A^0|0>\ =  f_{D_{1/2}} \ p'^3\ .
\eeq

\noi This is the convention that we have used in \cite{18r}, \cite{14r}
(there is a misprint in formula (14) of \cite{18r}, corrected in
\cite{14r}).\par

We find for the rates (only one helicity amplitude contributes to the
$B^- \to D_1^{1/2\ 0}\pi^-$ transition)~:
\bea
\label{33e}
&&\Gamma \left ( B^- \to D_0^{1/2\ 0} \pi^-\right ) \cong {G_F^2 \over 16 \pi} \left | V_{cb}V_{ud}^*\right |^2 {p \over m_B^2} \nn \\
&&\left [ a_1 \sqrt{m_Dm_B} \ 2\left ( m_B + m_D\right ) (w-1) f_{\pi} \tau_{1/2}(w) + a_2 \ m_B^2 \ f_{D_{1/2}} \ f_0^{\pi B} \left ( m_D^2\right ) \right ]^2
\eea
\bea
\label{34e}
&&\Gamma \left ( B^- \to D_1^{1/2\ 0} \pi^-\right ) \cong {G_F^2 \over 4 \pi} \left | V_{cb}V_{ud}^*\right |^2 {p \over m_B^2}\ {p^2 \over m_D^2} \nn \\
&&\left [ a_1 \sqrt{m_Dm_B} \left ( m_B - m_D\right ) f_{\pi} \tau_{1/2}(w) + a_2 \ m_B m_D \ f_{D_{1/2}} \ f_+^{\pi B} \left ( m_D^2\right ) \right ]^2 \ .
\eea

We will use in these expressions the color-allowed and color-suppressed
factors respectively of the order $a_1 \cong 1$, $a_2 \cong 0.3$ (B.8).
The powers of $p$ indicate that the decays $B \to D_0^{1/2}\pi$ and $B
\to D_1^{1/2}\pi$ occur respectively in the $S$ and $P$ waves. 
\par

The relative sign between both terms in (\ref{33e}) and (\ref{34e}) is
crucial. Let us give an argument that shows that the interference
is constructive. Instead of considering the $\pi$, let us
consider the pseudoscalar $D$ meson, composed of heavy-light quarks.
Our assumption is that the form factors and decay constants between
ground state mesons do not change sign when going from heavy mesons
made of heavy-light quarks to light mesons made of equal mass quarks. This is a
very sensible continuity hypothesis that is satisfied in the quark
model, since there are no nodes in these ground state wave functions, and the extrapolation in reduced mass is smooth. On the other
hand, this smooth continuity in mass is commonly used in lattice
calculations, and it is also observed, considering for example the decay constants $f_{D_s}$ or $f_K$ and varying the $c$ or $s$ quark masses.\par

In \cite{18r}, \cite{14r} we did demonstrate (using duality in
$B^0-\overline{B}^0$ mixing and also within the OPE) the following sum
rules in the heavy quark limit of QCD for heavy-light form factors and
decay constants, {\it valid for all values of $w$}~:
\beq
\label{35e}
\sum_n f^{(n)} \xi^{(n)}(w) = 2 \sum_n f_{1/2}^{(n)} \ \tau_{1/2}^{(n)}(w) = f^{(0)}
\eeq

\noi where $n$ denotes a radial quantum number, $f^{(0)}= f$ is the
ground state decay constant and $\xi^{(0)}(w) = \xi (w)$ the elastic
Isgur-Wise function. The decay constants $f^{(n)}$ and $f_{1/2}^{(n)}$
scale like ${1 \over \sqrt{m_Q}}$. In particular, we have demonstrated
that the rigorous SR (\ref{35e}) are satisfied within relativistic BT
quark models and in the non-relativistic quark model. Within BT quark
models, we have shown that a main contribution to the SR (\ref{35e})
comes from the $n = 0$ states, that has the same sign as the whole sum \cite{14r} and the same is true in the non-relativistic quark model~:
\beq
\label{36e}
{\rm Sign}  [f\ \xi (w)] = {\rm Sign} \left [ f_{1/2}\ \tau_{1/2}(w)\right ]
\eeq

\noi where we have used the notations $\tau_{1/2}^{(0)}(w) =
\tau_{1/2}(w)$, $f_{1/2}^{(0)} = f_{1/2}$. Multiplying the equalities
(\ref{36e}) by $\xi (w) \tau_{1/2}(w)$, we have Sign$\{f\tau_{1/2}(w)[\xi (w)]^2\} =$ Sign$\{f_{1/2}\xi
(w)$\break\noindent $[\tau_{1/2}(w)]^2\}$ and therefore 
\beq
\label{37e}
{\rm Sign} \left [ f \ \tau_{1/2}(w)\right ] = {\rm Sign} \left [f_{1/2}\ \xi (w) \right ]\ .
\eeq

\noi Heavy quark scaling implies for $B \to D$ form factors~:

\beq
\label{38e}
{\sqrt{4m_Dm_B} \over m_D + m_B}\ f_+^{DB}(q^2) = {\sqrt{4m_Dm_B} \over m_D + m_B}\ {f_0^{DB}(q^2)\over 1 - {q^2 \over (m_D + m_B)^2}}= \xi (w)
\eeq
\vskip 5 truemm

\noi and therefore Sign$[f_+^{DB}(q^2)] =$ Sign$[f_0^{DB}(q^2)] =$
Sign$[\xi (w)]$. Our continuum assumption linking heavy-light mesons to
light mesons implies then, within a definite phase convention~:
\bea
\label{39e}
&&{\rm Sign} \left [ f_+^{\pi B}(q^2)\right ] = {\rm Sign}\left [f_0^{\pi B}(q^2)\right ] = {\rm Sign}\left [ \xi (w)\right ]\nn \\
&&{\rm Sign} \left [ f_D\right ] = {\rm Sign}\left [f_{\pi}\right ]\ .
\eea

\noi From (\ref{37e}) and (\ref{39e}) we get
\beq
\label{40e}
{\rm Sign} \left [ f_{\pi }\tau_{1/2}(w)\right ] = {\rm Sign}\left [f_{1/2} \ f_+^{\pi B}(q^2)\right ] = {\rm Sign}\left [ f_{1/2}\ f_0^{\pi B} (q^2)\right ]\ . 
\eeq

\noi Therefore, a relative constructive sign between the two contributions
in (\ref{33e}) and (\ref{34e}) follows from (\ref{40e}). \par

We need some input on the form factors $f_0^{\pi B}(q^2)$ and $f_+^{\pi
B}(q^2)$. We could use the simple theoretically motivated pole-dipole
parametrization for $f_0^{\pi B}(q^2)$, $f_+^{\pi B}(q^2)$ of the Large
Energy Effective Theory (LEET) \cite{19r}~:
\beq
\label{41e}
f_0^{\pi B}(q^2) = \left ( 1 - {q^2 \over m_B^2}\right ) f_+^{\pi B}(q^2) \cong {0.3 \over 1 - {q^2 \over m_B^2}}\ .
\eeq

However, there is an empirical parametrization, inspired by (\ref{41e}), that fits the lattice
data on these form factors, proposed by Becirevic and Kaidalov
\cite{20r}~:
\bea
\label{42e}
&&f_+^{\pi B}(q^2) = {c_B \left ( 1 - \alpha_B\right ) \over \left ( 1 - {q^2 \over m_{B^*}^2}\right ) \left ( 1 - \alpha_B\  {q^2 \over m_{B^*}^2}\right )}\nn \\
&&f_0^{\pi B}(q^2) = {c_B \left ( 1 - \alpha_B \right ) \over \left ( 1 - {q^2 \over \beta_B\ m_{B(0^+)}^2}\right )}\ .
\eea

\noi A fit to the lattice data \cite{21r} yields two sets of values for
these parameters. We choose one of them, the other one yielding very
comparable results~:
\bea
\label{43e}
&&c_B = 0.51(8)(1)\nn \\
&&\alpha_B = 0.45(17)^{+ 0.06}_{- 0.13} \nn \\
&&\beta_B = 1.20(13)^{+ 0.15}_{- 0.00}
\eea

\noi that corresponds to
\beq
\label{44e}
f_+^{\pi B}(0) = f_0^{\pi B}(0) = 0.28(6)(5)\ .
\eeq

\noi Concerning the QCD coefficient $a_2$, in Appendix B we have made
an analysis of the well-measured decays $\overline{B} \to D(D^*)\pi$ in
all its charged modes. Since the perturbative estimation of $a_2 \cong
0.2$ \cite{22r} appears to give too small Class II branching ratios, we
have to consider, following \cite{2r}, \cite{3r} non-perturbative
contributions to $a_1$ and $a_2$ (see the discussion in Appendices B
and C), that yield the values
\beq
\label{45e}
a_1 \cong 1 \qquad \qquad \qquad a_2 \cong 0.3\ .
\eeq

\noi Moreover, we adopt, like in Appendix B, the parametrization for
the form factors $f_+^{\pi B}(q^2)$, $f_0^{\pi B}(q^2)$ given by
(\ref{42e})-(\ref{44e}). \par

Once we know the sign of the interference between the two terms in
(\ref{33e}) and (\ref{34e}), we proceed as follows. We extract the
decay constant $f_{D_{1/2}}$ from these formulas comparing to the Class I
ones
\beq
\label{46e}
\Gamma \left ( B^- \to D_0^{1/2\ 0} \pi^-\right ) = {G_F^2 \over 16 \pi} \left | V_{cb}V_{ud}^*\right |^2 {p \over m_B^2} 
\left [ a_1 \sqrt{m_Dm_B} \, 2\left ( m_B + m_D\right ) (w-1) f_{\pi}  \tau_{1/2}(w) \right ]^2
\eeq
\beq
\label{47e}
\Gamma \left ( B^- \to D_1^{1/2\ 0} \pi^-\right ) = {G_F^2 \over 4 \pi} \left | V_{cb}V_{ud}^*\right |^2 {p \over m_B^2}\ {p^2 \over m_D^2} 
\left [ a_1 \sqrt{m_Dm_B} \left ( m_B - m_D\right ) f_{\pi} \ \tau_{1/2}(w) \right ]^2\ .
\eeq

\noi Adding the theoretical errors in quadrature and using the QCD
coefficients (\ref{45e}), we find
\bea
\label{48e}
&f_{D_{1/2}} = (206 \pm 120)\ {\rm MeV} &\qquad\qquad ({\rm from}\ \overline{B} \to D_0^{1/2} \pi)\nn \\
&f_{D_{1/2}} = (196 \pm 93)\ {\rm MeV} &\qquad\qquad  ({\rm from}\ \overline{B} \to D_1^{1/2} \pi)\ .
\eea

\noi Both determinations are roughly consistent. In view of the large
systematic uncertainties, we proceed as in (\ref{18e}), taking the
union of both domains rather than the intersection. We thus keep the safe
range
\beq
\label{49e}
f_{D_{1/2}} = (206 \pm 120)\ {\rm MeV}\ .
\eeq

\subsection{Comparison with theoretical estimates of f$_{\bf D_{\bf 1/2}}$.}
\hspace*{\parindent} The value (\ref{49e}) is in reasonable
agreement with the calculation of QCDSR \cite{23r} that gives, for
decay constants of $D$ mesons with $0^-$ and $0^+$ quantum numbers,
including $1/m_Q$ and $\alpha_s$ corrections, the following numbers
\beq
\label{50e}
f_{D} = (195\pm 20) \ {\rm MeV} \qquad\qquad\qquad f_{D(0^+)} = (170\pm 20) \ {\rm MeV}\ .
\eeq

 There are also calculations within QCDSR in the heavy quark limit, without
including $\alpha_s$ corrections \cite{24r}, that give a larger value
for $f_{D(0^+)}$, consistent within $1\sigma$ with (\ref{49e})
\bea
\label{51e}
&\sqrt{m_D}\ f_{D} = (0.21 \pm 0.03) \ {\rm GeV}^{3/2}\nn \\
&\sqrt{m_{D(0^+)}}\ f_{D(0^+)} = (0.46 \pm 0.06) \ {\rm GeV}^{3/2}\ .
\eea

\noi Another estimation using QCDSR in the heavy quark limit \cite{25r} gives a larger value, 
\beq
\label{52e}
\sqrt{m_{D(0^+)}}\ f_{D(0^+)} = (0.570 \pm 0.08)\  {\rm GeV}^{3/2}
\eeq

\noi and correcting for the $B\pi$ continuum \cite{25r} one gets the results
\bea
\label{53e}
&\sqrt{m_D}\ f_{D} = 0.35 \ {\rm GeV}^{3/2}\nn \\
&\sqrt{m_{D(0^+)}}\ f_{D(0^+)} = (0.36 \pm 0.10) \ {\rm GeV}^{3/2}
\eea

\noi that are consistent with (\ref{49e}). \par

Within the Bakamjian-Thomas class of relativistic quark models
\cite{10r}, the decay constants of heavy-light mesons in the heavy quark
limit have been computed \cite{14r}. One finds heavy quark scaling
$\sqrt{m_D}f_D =$ Const. for the $D(D^*)$ states and for the
doublet of $j^P = {1\over 2}^+$ states. Within the specific
spectroscopy model of Godfrey and Isgur \cite{27r}, one finds the
following values for the lowest $n = 0$ states~:
\bea
\label{54e}
&&\sqrt{m_D}\ f_D = (0.670 \pm 0.020) \ {\rm GeV}^{3/2} \nn \\
&&\sqrt{m_{D_{1/2}}}\ f_{D_{1/2}} = (0.640 \pm 0.020) \ {\rm GeV}^{3/2}\ .
\eea

These values for ${1 \over 2}^-$ and ${1 \over
2}^+$ doublets are close. From the masses $m_{D(D^*)}$ and $m_{D_{1/2}}$ one finds,
in the heavy quark limit
\bea
\label{55e}
&&f_D \cong f_{D^*} \cong (474 \pm 14)\ {\rm MeV} \\
&&f_{D_{1/2}} \cong (417 \pm 13)\ {\rm MeV}\ .
\label{56e}
\eea

\noi These values for $f_D$, $f_{D^*}$ are {\it much larger} than the
estimations given by lattice QCD \cite{28r}, \cite{29r}, even adding a 10 \%
error due to quenching (used in Appendix B)~:
\bea
\label{57e}
&&f_D = (216 \pm 36)\ {\rm MeV}\nn \\
&&f_{D^*} = (258 \pm 52)\ {\rm MeV} \ .
\eea

\noi For $f_{D(0^+)}$ one finds, in lattice QCD, keeping only the statistical error \cite{33newreference},
\beq
\label{59e}
f_{D(0^+)} = (122 \pm 43)\ {\rm MeV} \ .
\eeq

\noi This latter value is consistent with the value obtained in the quark model of Veseli and Dunietz \cite{19newreference}~: 

\beq
\label{59newe}
f_{D(0^+)} = (139 \pm 30)\ {\rm MeV} \ .
\eeq

These theoretical estimations of the $f_{D_{1/2}}$ or $f_{D(0^+)}$
decay constants, that become equal in the heavy quark limit, are not
homogeneous in their methods. To make the panorama somewhat clearer, we
summarize the results in Tables 2 and 3. In Table 2 we give the results
of the different methods {\it at finite mass}, together with the
phenomenological determination of the present paper. In Table 3 we give
the results of the methods in the heavy quark limit, dividing the
invariant $\sqrt{m_Q} f_{1/2}$ by $\sqrt{m_{D(0^+)}}$ with $m_{D(0^+)}
= 2290$~MeV of the Belle experiment (Appendix A). Although, of course,
this choice is somewhat arbitrary, one can thus qualitatively compare
with the finite mass results.

\begin{center}
\begin{tabular}{|c|c|}
\hline
Theoretical method &$f_{D(0^+)}$ \\
\hline
QCD Sum Rules  \cite{23r} &$(170 \pm 20)$ MeV\\
\hline
Lattice QCD   \cite{33newreference} &$(122 \pm 43)$ MeV\\
  \hline
 Veseli-Dunietz quark model \cite{19newreference} &$(139 \pm 30)$ MeV\\
\hline
 Present phenomenological &$(206 \pm 120)$ MeV\\
 determination from $B \to D^{**}\pi$ & \\
 \hline
\end{tabular}
\end{center}
\noi {\bf Table 2.} Theoretical predictions for the decay constant
$f_{D(0^+)}$ in the different methods,  {\it at finite mass}, compared with the phenomenological determination of the
present paper.\par \vskip 1.5 truecm

\begin{center}
\begin{tabular}{|c|c|}
\hline
Theoretical method &${\sqrt{m_Q}\ f_{1/2} \over \sqrt{m_{D(0^+)}}}$ \\
\hline
QCD Sum Rules  \cite{24r} &$(304 \pm 40)$ MeV\\
\hline
QCD Sum Rules  \cite{25r} &$(377 \pm 53)$ MeV\\
\hline
QCD Sum Rules &$(238 \pm 66)$ MeV\\
 (correcting for $B\pi$ continuum) \cite{25r} &\\
 \hline
 Bakamjian-Thomas quark model  \cite{14r} &$(417 \pm 13)$ MeV\\
 \hline
\end{tabular}
\end{center}
\noi {\bf Table 3.} Theoretical predictions for $\sqrt{m_Q} \ f_{1/2}$ {\it in the heavy quark limit}, divided by $\sqrt{m_{D(0^+)}}$, with $m_{D(0^+)}
= 2290$~MeV of the Belle experiment.\par \vskip 5 truemm

The table shows a very scattered set of results, but there is the general trend that the decay constant is much larger in the
methods that use the heavy quark limit. The largest value is obtained by the BT models. The subleading correction is negative. Also, we observe that the phenomenological
determination of the present paper, that has a large error, agrees within errors
with the methods including finite mass corrections.

\section{Predictions for Class II decays.} \hspace*{\parindent}
Let us finish our discussion giving
predictions for the rates of the color suppressed
decays, using the range (\ref{49e}) for the decay constant $f_{D_{1/2}}$.
From the rates 
\bea
\label{63e}
&&\Gamma \left ( \overline{B}^0 \to D_0^{1/2\ 0} \pi^0\right ) = {G_F^2 \over 16 \pi} \left | V_{cb}V_{ud}^*\right |^2 {p \over m_B^2} 
{1\over 2}\left [ a_2 m_B^2 f_{D_{1/2}} f_0^{\pi B} (m_D^2) \right ]^2\nn \\
&&\Gamma \left ( \overline{B}^0 \to D_1^{1/2\ 0} \pi^0\right ) = {G_F^2 \over 4 \pi} \left | V_{cb}V_{ud}^*\right |^2 {p \over m_B^2} 
 {p^2 \over m_D^2} {1\over 2}\left [ a_2 m_B m_D f_{D_{1/2}} f_+^{\pi B} (m_D^2) \right ]^2
\eea

\noi we obtain the branching ratios
\bea
\label{64e}
&&BR\left ( \overline{B}^0 \to D_0^{1/2\ 0} \pi^0\right ) = (2.8 \pm 2.0) \times 10^{-4}\nn \\
&&BR\left ( \overline{B}^0 \to D_1^{1/2\ 0} \pi^0\right ) = (2.2 \pm 1.5) \times 10^{-4}\ .
\eea

\noi The central values are large enough that could in principle be
measured. These rates are independent of the IW function $\tau_{1/2}(w)$,
while they depend on the non-vanishing decay constant $f_{D_{1/2}}$.\par

Heavy quark symmetry plus factorization predicts
\beq
\label{65e}
BR\left ( \overline{B}^0 \to D_2^{3/2\ 0} \pi^0\right ) = BR\left ( \overline{B}^0 \to D_1^{3/20} \pi^0\right ) = 0 
\eeq

\noi because of the vanishing of the $f_{3/2}$ decay
constants (\ref{27e}). However, it is worth noticing that, because of the large
experimental errors and theoretical uncertainties (spin counting,
etc.), from the BR (\ref{26e}) and using the isospin relation
(\ref{1e}) we can only have a rather loose upper bound
\bea
\label{66e}
&&BR\left ( \overline{B}^0 \to D_2^{3/2\ 0} \pi^0\right ) < 4 \times 10^{-4} \nn \\
&&BR\left ( \overline{B}^0 \to D_1^{3/2\ 0} \pi^0\right ) < 3 \times 10^{-4}\ .
\eea

\section{The rate to excited states in semileptonic B\break\noindent decays.} \hspace*{\parindent} 
The values of the functions $\tau_{1/2}(w)$, $\tau_{3/2}(w)$ at $w= 1$ and
their $w$-dependence gives predictions for the
semileptonic (SL) decay $\overline{B} \to D^{**}\ell \nu$ branching
ratios in the heavy quark limit. The differential decay rates for $\overline{B} \to
D_J^j$ $(j = {1 \over 2}, {3 \over 2})\ell \nu$ write \cite{14r}
\bea
\label{23e}
{d\Gamma \left ( B \to D_2^{3/2}\ell \nu \right ) \over dw} &=& {G_F^2 m_B^5 \over 48 \pi^3} \left |V_{cb}\right |^2 2r^3 (w+1) (w^2 - 1)^{3/2}\nn \\
&&\left [ (w+1) (1-r)^2 + 3w(1+r^2-2rw)\right ] \left | \tau_{3/2}(w)\right |^2\nn \\
&& \nn \\
{d\Gamma \left ( B \to D_1^{3/2}\ell \nu \right ) \over dw} &=& {G_F^2 m_B^5 \over 48 \pi^3} \left |V_{cb}\right |^2 2r^3 (w+1) (w^2 - 1)^{3/2}\nn \\
&&\left [ (w-1) (1+r)^2 + w(1+r^2-2rw)\right ] \left | \tau_{3/2}(w)\right |^2\nn \\
&&\nn\\
{d\Gamma \left ( B \to D_1^{1/2}\ell \nu \right ) \over dw} &=& {G_F^2 m_B^5 \over 48 \pi^3} \left |V_{cb}\right |^2 4r^3 (w-1) (w^2 - 1)^{1/2}\nn \\
&&\left [ (w-1) (1+r)^2 + 4w(1+r^2-2rw)\right ] \left | \tau_{1/2}(w)\right |^2
\eea
$${d\Gamma \left ( B \to D_0^{1/2}\ell \nu \right ) \over dw} = {G_F^2 m_B^5 \over 48 \pi^3} \left |V_{cb}\right |^2 4r^3 (w^2 - 1)^{3/2}
(1-r)^2 \left | \tau_{1/2}(w)\right |^2$$
\vskip 5 truemm

\noi where $r = {m_D \over m_B}$ and $D$ denotes the corresponding
$D_J^j$ meson. \par

For completeness, we write down the corrresponding formulas for the ground state~:
$${d\Gamma (B \to D\ell \nu ) \over dw} = {G_F^2 m_B^5 \over 48 \pi^3} |V_{cb}|^2  r^3 (w^2 - 1)^{3/2} (1 + r)^2 |\xi (w) |^2$$
\bea
\label{equation67}
{d\Gamma (B \to D^*\ell \nu )\over dw} &=& {G_F^2 m_B^5 \over 48 \pi^3} |V_{cb}|^2 r^{3} (1+w) (w^2 - 1)^{1/2}\nn \\
&&\left [ (w+1) (1 - r)^2 + 4w (1 + r^{2} - 2rw)\right ] |\xi (w)|^2 \ .
\eea

\noi The situation is given in Table 4, a slight modification of the predictions of ref. \cite{14r}.\par \vskip 5 truemm

\begin{tabular}{|c|c|c|}
\hline
Semileptonic mode &Experiment &Model\\
\hline
$B \to D \ell \nu$ &$(2.14 \pm 0.20) \times 10^{-2}$ &$(1.95 \pm 0.45) \times 10^{-2}$\\
\hline
$B \to D^* \ell \nu$ &$(5.44 \pm 0.23) \times 10^{-2}$ &$(5.90 \pm 1.10) \times 10^{-2}$\\
\hline
&(a) $(2.4 \pm 1.1) \times 10^{-3}$ &\\
$B \to D_2^{3/2} \ell \nu$ &(b) $(4.4 \pm 2.4) \times 10^{-3}$ &$\left ( 6.3^{+ 3.0}_{ - 2.0}\right ) \times 10^{-3}$\\
&(c) $(3.0 \pm 3.4) \times 10^{-3}$ & \\
\hline
&(a) $(7.0 \pm 1.6) \times 10^{-3}$ & \\
$B \to D_1^{3/2}\ell \nu$ &(b) $(6.7 \pm 2.1) \times 10^{-3}$ &$\left ( 4.0^{+ 1.2}_{- 1.4}\right ) \times 10^{-3}$\\
&(c) $(5.6 \pm 1.6) \times 10^{-3}$ &  \\
\hline
$B \to D_1^{1/2} \ell \nu$ &(b) $(2.3 \pm 0.7) \times 10^{-2}$ &$(6 \pm 2) \times 10^{-4}$ \\
\cline{1-1}  \cline{3-3}
$B \to D_0^{1/2}\ell \nu$ &``wide'' $D^{**} \to (D + D^*)\pi$ &$(6\pm 2) \times 10^{-4}$\\
\hline
\end{tabular}

\vskip 3 truemm

\noi {\bf Table 4.} Comparison between rates for $\overline{B} \to
D_J^j\ell \nu$ decays and the model described in the text. The data are
from (a) ALEPH \cite{reference32} \cite{15r}, (b) DELPHI \cite{15r}
\cite{17newreference} \cite{reference33} and (c) CLEO \cite{16r}
experiments. For the elastic IW function we adopt the values (B.5),
(B.6) of Appendix B. ``Wide'' stands for unidentified $(D + D^*)\pi$
events forming a wide bump. In the text we discuss a new DELPHI analysis \cite{nouvelleref} and very recent data from the Belle collaboration \cite{neo}.\par \vskip 5 truemm

To make predictions for the SL rates we need an input on the IW
functions $\tau_{1/2}(w)$, $\tau_{3/2}(w)$. First, we must take into
account that it is reasonable to expect that the $n = 0$ IW functions
give a sizeable contribution to Bjorken and Uraltsev SR. Making this
assumption, Uraltsev SR is very constraining on the difference
$|\tau_{3/2}(1)|^2 - |\tau_{1/2}(1)|^2$, that should be not far away
from ${1 \over 4}$. This is the case for the BT model values
(\ref{21e}), and, more importantly, consistent with the values that we
have found from non-leptonic decays (\ref{20e}) using the Belle data.
As an example, we will then adopt the values $|\tau_{3/2}(1)|$ and
$|\tau_{1/2}(1)|$ given by the BT model (\ref{21e}) and allow
nevertheless a $\pm 50 \%$ departure for the values of the slopes. The
lower values of the slopes would be in agreement with ref.
\cite{26newref}. This gives the range of model predictions for SL rates
in the table. Although the $B \to D_J^{3/2}\ell \nu$ rates are in
reasonable agreement, for the $B \to D_J^{1/2}\ell \nu$ rates there is
a problem. We find $BR[\overline{B} \to (D_0^{1/2} + D_1^{1/2})\ell \nu
] = (1.2 \pm 0.4) \times 10^{-3}$. The DELPHI experiment gives, once
the $BR [\overline{B} \to (D_2^{3/2} + D_1^{3/2})\ell \nu]$ is
subtracted, a large branching ratio of $B$ decays into ``wide''
$D^{**}$ mesons decaying into $(D + D^*)\pi$. We will call this
branching ratio $BR_{wide}^{DELPHI}[\overline{B} \to (D + D^*)\pi \ell
\nu ] = (2.3 \pm 0.7) \times 10^{-2}$, that we report in the last line
of Table 4. On the one hand, this BR is one order of magnitude larger
than our prediction for $BR [\overline{B} \to (D_0^{1/2} +
D_1^{1/2})\ell \nu]$. On the other hand, keeping only to the DELPHI
experiment, one obtains the sum  $BR[\overline{B} \to (D + D^*)\ell
\nu] + BR [\overline{B} \to (D_2^{3/2} + D_1^{3/2})\ell \nu ] +
BR_{wide} [\overline{B} \to (D + D^*)\pi \ell \nu ] = (11 \pm 1.6)
\times 10^{-2}$, that already saturates within errors, that are however
large, the total semileptonic width $BR [\overline{B} \to  \ell \nu +
{\rm anything}] = (10.73 \pm 0.28) \times 10^{-2}$. There are therefore
two problems. On the one hand, other wide states besides $D_0^{1/2} +
D_1^{1/2}$ have to contribute to $BR_{wide}[\overline{B} \to (D +
D^*)\pi \ell \nu]$. These could be radial excitations or higher orbital
excitations, that are in principle allowed due to the large phase space
available. The experimental width is so large that it includes high
masses. Moreover, the multiplicity of higher excitations grows with the
mass. On the other hand, it is curious that considering only the modes
$\overline{B} \to (D + D^*)\ell \nu$, $\overline{B} \to (D + D^*) \pi
\ell \nu$ the total semileptonic width is already saturated, and one
could wonder why there is no place for decays into multipion modes
$\overline{B} \to (D + D^*) + n \pi$ ($n > 1$) and why they are not
observed. Presumably, due to phase space, these could not come from
modes of the type $D\rho$ but could come from various $D^{**}\pi$. \par

A recent new analysis by DELPHI \cite{nouvelleref} confirms and makes
more precise this situation. The total branching ratio into $D^{**}$
(narrow and broad) is measured to be
\beq
\label{Reference66e}
BR \left ( \overline{B}^0 \to D^{**}\ell \overline{\nu}\right ) = (2.7 \pm 0.7 \pm 0.2 ) \%
\eeq

\noi with the decay final states dominated by the $D(D^*)\pi$ channels.
The dominant contributing channel is a broad state decaying into
$D^*\pi$, i.e. a state $D^{**}$ or $J^P = 1^+$ with a mass $M = 2445
\pm 34 \pm 10$~MeV and a width $\Gamma = 234 \pm 74 \pm 25$~MeV. On the
other hand, broad $D\pi$ states favor a production with a maximum close
to threshold. On the other hand, DELPHI bounds the branching ratios
into $D\pi\pi$ and $D^*\pi\pi$ final states.\par

On the other hand, very interesting recent data have been published by
Belle at the last Lepton-Photon Conference \cite{neo}, that gives the
following branching ratios
\bea
\label{Reference67e}
&&BR \left ( B^- \to D^+\pi^-\ell^-\overline{\nu}\right ) = (0.54 \pm 0.07 \pm 0.07 \pm 0.06) \times 10^{-2}\nn \\
&&BR \left ( B^- \to D^{*+}\pi^-\ell^-\overline{\nu}\right ) = (0.67 \pm 0.11 \pm 0.09 \pm 0.03) \times 10^{-2}\nn \\
&&BR \left ( \overline{B}^0 \to D^0\pi^+\ell^-\overline{\nu}\right ) = (0.33 \pm 0.06 \pm 0.06 \pm 0.03) \times 10^{-2}\nn \\
&&BR \left ( \overline{B}^0 \to D^{*0}\pi^+\ell^-\overline{\nu}\right ) = (0.65 \pm 0.12 \pm 0.08 \pm 0.05) \times 10^{-2} \ .
\eea

\noi These values are to be compared with the theoretical expectations
of Table 4. We find, considering only the central values of the model
and taking into account the relevant branching fractions of the
different $D_J^j$ computed in Section 2,
\bea
\label{Reference68e}
&&BR \left ( B^- \to D^+\pi^-\ell^-\overline{\nu}\right ) = BR \left ( \overline{B}^0 \to D^{0}\pi^+\ell^-\overline{\nu}\right ) = 0.43 \times 10^{-2}\nn \\
&&BR \left ( B^- \to D^{*+}\pi^-\ell^-\overline{\nu}\right ) = BR \left ( \overline{B}^0 \to D^{*0}\pi^+\ell^-\overline{\nu}\right ) = 0.34 \times 10^{-2}
\eea

In view of the uncertainties, there is a fair agreement between our predictions and the Belle data.

In conclusion, there seems to be a potential problem concerning the DELPHI semi-leptonic
data on $B\to (D_1^{1/2} + D_0^{1/2})\ell \nu$ decays that should be
addressed in future experiments. If $BR_{wide} [B \to (D + D^*)\pi \ell
\nu]$ had to be attributed to $D_1^{1/2} + D_0^{1/2}$, then
$\tau_{1/2}(1)$ would be much larger than $\tau_{3/2}(1)$, in
contradiction with the expectations of Uraltsev SR. This is at odds with  
the Belle non-leptonic data studied in the present paper.
The study of the $D_J^j$ wide states is not an easy experimental task.
A recent Tevatron D0 experiment sees clearly in SL B decays the narrow
states $j = {3 \over 2}$ but has not given a measurement of the wide
ones $j = {1 \over 2}$ \cite{17r}. On the other hand, we find agreement with the very recent Belle data on $\overline{B} \to D(D^*)\pi \ell \nu$.

\section{Conclusion.} \hspace*{\parindent}
In conclusion, we have shown within a simple factorization model,
tested in the well-measured $B \to D(D^*)\pi$ decays, that one can extract information on the
Isgur-Wise functions at zero recoil $\tau_{1/2}(1)$, $\tau_{3/2}(1)$
from non-leptonic data on $\overline{B}^0 \to D^{**+} \pi^-$ (Class
I). Combining with $B^- \to D^{**0}\pi^-$ (Class III) one can
obtain the non-vanishing decay
constant $f_{D_{1/2}}$ of $D^{**}(j = {1\over 2})$. Special care has been taken in the
determination of the interference sign between the $\pi$ and $D^{**}$
emission diagrams. The ranges obtained for $\tau_{1/2}(1)$,
$\tau_{3/2}(1)$ are consistent for both types of modes, with the
expectations of Bjorken and Uraltsev sum rules, and with the predictions
of the Bakamjian-Thomas quark model of form factors. Moreover, the
range of values found for the decay constant $f_{D_{1/2}}$ agree with most
theoretical expectations. We predict sizeable rates of the Class II
decays $\overline{B}^0 \to D^{**0} \pi^0$ that could be measured in
the near future. On the contrary, for $D^{**}(j = {3 \over 2})$,
Class II decays should be suppressed due to the vanishing of the decay
constant $f_{3/2}$ in the heavy quark limit. \par

We must warn that $1/m_Q$ corrections could be large and upset the
results of the present stage of this analysis, as discussed in Appendix
D. Also, we point out a problem with present DELPHI data of
semileptonic decays for the total rate to excited states. Very recent
Belle data on $\overline{B} \to D^{(*)}\pi\ell\nu$ seem in good
agreement with the theoretical expectations.

\section*{Acknowledgements}
\hspace*{\parindent}
We are indebted to A. Bondar from the Belle Collaboration that pointed out
to us that the $D^{**0}$ emission diagram could be very important in
Class III decays. We acknowledge also D. Be\'cirevi\'c for providing unpublished lattice results on $f_{D(0^+)}$, and I. Bigi for recalling to us semileptonic
data from the D0 Collaboration. We are also indebted to the EC contract
HPRN-CT-2002-00311 (EURIDICE) for its support.

\section*{Appendix A. Belle data on B $\to$ D$^{\bf **}\pi$ decays.} \hspace*{\parindent}
For the sake of clarity on how we extract the branching ratios of the different $B \to D^{**}\pi$ decay modes, and how we handle the errors in the text, we
reproduce here the Belle data on Class I \cite{5r} and Class
III decays \cite{4r}.

\begin{center}
 \begin{tabular}{|c|c|} 
\hline
$M_2^{3/2}$ &$\left (2459.5 \pm 2.3 \pm 0.7^{+4.9}_{- 0.5}\right )$ MeV\\
\hline
$\Gamma_2^{3/2}$ &$(48.9 \pm 5.4 \pm 4.2 \pm 1.0)$ MeV\\
\hline
$B\left ( \overline{B}^0 \to D_2^{3/2\ +} \pi^-\right ) B\left ( D_2^{3/2\ +} \to D^0\pi^+\right )$ &$\left (3.08 \pm 0.33 \pm 0.09^{+0.15}_{- 0.02}\right )\times  10^{-4}$ \\
\hline
$B\left ( \overline{B}^0 \to D_2^{3/2\ +} \pi^-\right ) B\left ( D_2^{3/2\ +} \to D^{*0}\pi^+\right )$ &$\left (2.45 \pm 0.42^{+0.35\  + 0.39}_{- 0.45 \ - 0.17}\right )\times  10^{-4}$ \\
\hline
$M_1^{3/2}$ &$(2428.2\pm 2.9 \pm 1.6 \pm 0.6)$ MeV\\
\hline
$\Gamma_1^{3/2}$ &$\left (34.9 \pm 6.6^{+4.1}_{- 0.9}\pm 4.1\right )$ MeV\\
\hline
$B\left ( \overline{B}^0 \to D_1^{3/2\ +} \pi^-\right ) B\left ( D_1^{3/2\ +} \to D^{*0}\pi^+\right )$ &$\left (3.68 \pm 0.60^{+0.71 \ + 0.65}_{- 0.40 \ - 0.30}\right )\times  10^{-4}$ \\
\hline
$M_0^{1/2}$ &$(2290\pm 22 \pm 20)$ MeV\\
\hline
$\Gamma_0^{1/2}$ &$(26.7 \pm 3.1 \pm 2.2)$ MeV\\
\hline
$B\left ( \overline{B}^0 \to D_0^{1/2\ +} \pi^-\right ) B\left ( D_0^{1/2\ +} \to D^0\pi^+\right )$ &$< 1.2 \times  10^{-4}$ \\
\hline
$M_1^{1/2}$ &$(2428 \pm 2.9 \pm 1.6 \pm 0.6)$ MeV\\
\hline
$\Gamma_1^{1/2}$ &$(380\pm 100\pm 100)$ MeV\\
\hline
$B\left ( \overline{B}^0 \to D_1^{1/2\ +} \pi^-\right ) B\left ( D_1^{1/2\ +} \to D^{*0}\pi^+\right )$ &$< 0.7 \times  10^{-4}$ \\
\hline
\end{tabular} 
\end{center}
\noi {\bf Table 5.}
Data of the Belle Collaboration for the masses, widths and branching
ratios to $D^{**}$ states $D_J^j$ for Class I decays $\overline{B}^0 \to D^{**+}\pi^-$
\cite{5r}.\par

\begin{center} \begin{tabular}{|c|c|} 
\hline
$M_2^{3/2}$ &$(2461.6 \pm 2.1 \pm 0.5 \pm 3.3)$ MeV\\
\hline
$\Gamma_2^{3/2}$ &$(45.6 \pm 4.4 \pm 6.5 \pm 1.6)$ MeV\\
\hline
$B\left ( {B}^- \to D_2^{3/2\ 0} \pi^-\right ) B\left ( D_2^{3/2\ 0} \to D^+\pi^-\right )$ &$(3.4 \pm 0.3 \pm 0.6 \pm 0.4)\times  10^{-4}$ \\
\hline
$B\left ( {B}^- \to D_2^{3/2\ 0} \pi^-\right ) B\left ( D_2^{3/2\ 0} \to D^{*+}\pi^-\right )$ &$(1.8 \pm 0.3 \pm 0.3 \pm 0.2)\times  10^{-4}$ \\
\hline
$M_1^{3/2}$ &$(2421.4\pm 1.5 \pm 0.4 \pm 0.8)$ MeV\\
\hline
$\Gamma_1^{3/2}$ &$(23.7 \pm 2.7 \pm 0.2\pm 4.0)$ MeV\\
\hline
$B\left ( {B}^- \to D_1^{3/2\ 0} \pi^-\right ) B\left ( D_1^{3/2\ 0} \to D^{*+}\pi^-\right )$ &$(6.8 \pm 0.7 \pm 1.3 \pm 0.3)\times  10^{-4}$ \\
\hline
$M_0^{1/2}$ &$(2308\pm 17 \pm 15 \pm 28)$ MeV\\
\hline
$\Gamma_0^{1/2}$ &$(276 \pm 21 \pm 18 \pm 60)$ MeV\\
\hline
$B\left ( {B}^- \to D_0^{1/2\ 0} \pi^-\right ) B\left ( D_0^{1/2\ 0} \to D^+\pi^-\right )$ &$(6.1 \pm 0.6 \pm 0.9 \pm 1.6)\times  10^{-4}$ \\
\hline
$M_1^{1/2}$ &$(2427.0 \pm 26 \pm 20 \pm 15)$ MeV\\
\hline
$\Gamma_1^{1/2}$ &$\left ( 384^{+107}_{- 75}\pm 24\pm 70\right )$ MeV\\
\hline
$B\left ( {B}^- \to D_1^{1/2\ 0} \pi^-\right ) B\left ( D_1^{1/2\ 0} \to D^{*+}\pi^-\right )$ &$(5.0\pm 0.4 \pm 1.0 \pm 0.4) \times  10^{-4}$ \\
\hline
\end{tabular} 
\end{center}
\noi {\bf Table 6.}
Data of the Belle Collaboration for the masses, widths and branching
ratios to $D^{**}$ states $D_J^j$ for Class III decays $B^- \to D^{**0}\pi^-$
\cite{4r}.\par \vskip 5 truemm

\section*{Appendix B. Testing the factorization model in\break\noindent B
$\to$ D(D$^{\bf *}$)$\pi$.}\hspace*{\parindent}
In this Appendix, in order to check qualitatively our simple
factorization model applied to $B \to D^{**}\pi$ decay, we use it to
describe the well measured decays into the ground state $B \to
D(D^*)\pi$. The data of the PDG \cite{12r} is given in Table 6,
together with our predictions that follow from the following simple
formulae. \par

From the definitions
$$<\pi^-(q)|A_{\mu}|0>\ = f_{\pi}\ q_{\mu}$$
$$<D^0(p')|A_{\mu}|B^-(p)>\ = \sqrt{m_Dm_B} \ \xi(w)\ (v + v')_{\mu}$$
$$<D^{*0}(p', \varepsilon ) |V_{\mu} |B^-(p)>\ = \sqrt{m_Dm_B}\ \xi (w) \left [ (w+1) \varepsilon_{\mu}^* - (\varepsilon^* \cdot v) v'_{\mu}\right ]$$
$$<D^0(p')|A_{\mu}|0>\ = f_D\ p'_{\mu}$$
$$<D^{*0}(p', \varepsilon )|V_{\mu}|0> \ = f_D m_{D^*}\ \varepsilon_{\mu}^*$$
$$<\pi^-(q)|V_{\mu}|B^-(p)>\ = \left ( p_{\mu} + q_{\mu} - {m_B^2 - m_{\pi}^2 \over p'^2} p'_{\mu}\right ) f_+^{\pi B}(p'^2) + {m_B^2 - m_{\pi}^2 \over p'^2} p'_{\mu}\ f_0^{\pi B}(p'^2)\eqno({\rm B.1})$$

\noi we obtain the matrix elements, satisfying the isospin relation (\ref{1e}), 
$$A\left ( B^- \to D^0\pi^-\right ) = a_1 \sqrt{m_Dm_B} \left ( m_B - m_D\right ) (w+1) f_{
\pi}\xi (w) + a_2 \ m_B^2\ f_D \ f_0^{\pi B}(m_D^2)$$
$$A\left ( \overline{B}^0 \to D^+\pi^-\right ) = a_1 \sqrt{m_Dm_B} \left ( m_B - m_D\right ) (w+1) f_{\pi} \xi (w)$$
$$A\left ( \overline{B}^0 \to D^0\pi^0\right ) = - \ a_2 \ {1 \over \sqrt{2}} \ m_B^2\ f_D\ f_0^{\pi B}(m_D^2) \eqno({\rm B.2})$$

$$A\left ( B^- \to D^{*0}\pi^-\right ) = {p \over m_{D^*}} \Big [ a_1 \sqrt{m_{D^*}m_B} \left ( m_B + m_{D^*}\right )  
f_{\pi}\xi (w)$$ 
$$+ \ a_2 \ 2m_{D^*}m_B\ f_{D^*} \ f_+^{\pi B}(m_{D^*}^2)\Big ]$$
$$A\left ( \overline{B}^0 \to D^{*+}\pi^-\right ) = a_1 \ {p \over m_{D^*}}\ \sqrt{m_{D^*}m_B} \left ( m_B + m_{D^*}\right ) f_{\pi} \xi (w)$$
$$A\left ( \overline{B}^0 \to D^{*0}\pi^0\right ) =  - \ a_2 \ {1 \over \sqrt{2}} \ {p \over m_{D^*}} \ 2m_{D^*} m_B\ f_{D^*}\ f_+^{\pi B}(m_{D^*}^2) \ .\eqno({\rm B.3})$$

\vskip 5 truemm
\noi For the decay constants $f_D$, $f_{D^*}$ we use the values of
lattice calculations within the quenched aproximation $f_D =
216(11)(5)$~MeV \cite{28r}, $f_{D^*} = 258(14)(6)$~MeV \cite{29r},
where the first error is statistical and the second is systematic.
Assuming a 10~\% uncertainty
due to the quenching approximation and adding all the errors in quadrature, we adopt the values
$$f_D = (0.216 \pm 0.036)\ {\rm GeV}$$
$$f_{D^*} = (0.258 \pm 0.052)\ {\rm GeV} \eqno({\rm B.4})$$

\noi For the form factors $f_0^{\pi B}(q^2)$, $f_+^{\pi B}(q^2)$ we use
(\ref{42e})-(\ref{44e}), and for the Isgur-Wise function, we use the
parametrization given by the BT model (last reference of \cite{10r}),
$$\xi (w) \cong \left ( {2 \over w+1}\right )^{2\rho^2} \eqno({\rm B.5})$$

\noi that we have used in \cite{30r} to fit Belle data on $B \to
D^*\ell \nu$ \cite{31r}, that gives ${\cal F}^*(1) |V_{cb}| = 0.036 \pm
0.002$ and $\rho^2 = 1.15 \pm 0.18$. 
In conclusion, from ${\cal F}^*(1) \cong 0.91$, we adopt the ranges
$$|V_{cb}| = 0.040 \pm 0.002$$
$$\rho^2 = 1.15 \pm 0.18 \ . \eqno({\rm B.6})$$

\noi We add the theoretical errors in quadrature, that gives, in
amplitude, a 10 \% error for Class I decays and a 30 \% error for Class
II decays. Adopting the values
$$a_1 \cong 1\ , \qquad a_2 \cong 0.2 \eqno({\rm B.7})$$

\noi obtained by the perturbative calculations \cite{22r}, and
considering the uncertainties given in (B.4) and (B.6), we obtain the
predictions of Table 7.\par \vskip 5 truemm

\newpage
\begin{center}
\begin{tabular}{|c|c|c|}
\hline
Modes &Experiment &Factorization\\
$B \to D(D^*)\pi$ & &$a_1 \cong 1$, $a_2 \cong 0.2$\\
\hline
$BR \left ( \overline{B}^0 \to D^+\pi^-\right )$ &$(2.76 \pm 0.25) \times 10^{-3}$ &$(3.4 \pm 0.7) \times 10^{-3}$\\
\hline
$BR \left ( \overline{B}^0 \to D^0\pi^0\right )$ &$(2.7 \pm 0.8) \times 10^{-4}$ &$(0.6 \pm 0.4) \times 10^{-4}$\\
\hline
$BR \left ( B^- \to D^0\pi^-\right )$ &$(4.98 \pm 0.29) \times 10^{-3}$ &$(5.2 \pm 1.0) \times 10^{-3}$\\
\hline
$BR \left ( \overline{B}^0 \to D^{*+}\pi^-\right )$ &$(2.76 \pm 0.21) \times 10^{-3}$ &$(3.3 \pm 0.7) \times 10^{-3}$\\
\hline
$BR \left ( \overline{B}^0 \to D^{*0}\pi^0\right )$ &$(2.7 \pm 0.5) \times 10^{-4}$ &$(0.8 \pm 0.5) \times 10^{-4}$\\
\hline
$BR \left ( B^- \to D^{*0}\pi^-\right )$ &$(4.6 \pm 0.4) \times 10^{-3}$ &$(5.3 \pm 1.0) \times 10^{-3}$\\
\hline
\end{tabular}
\end{center}

\noi {\bf Table 7.} Data on the branching ratios of $B \to D(D^*)\pi$
from PDG 2004 \cite{12r}, and the predictions of the factorization
model with the perturbative values $a_1 \cong 1$, $a_2 \cong 0.2$. The errors in the
predictions come from the uncertainty on the decay constants of
$D(D^*)$ mesons, the value of $|V_{cb}|$, the Isgur-Wise
function and the form factors $f_+^{\pi B}(q^2)$, $f_0^{\pi
B}(q^2)$ (\ref{42e})-(\ref{44e}). The theoretical errors are added in
quadrature. \par \vskip 5 truemm

\noi A first remark on the experimental data of Table 7 is that the
isospin relation (\ref{1e}) is roughly satisfied by real amplitudes,
i.e. without the need of FSI phases, like in the naive factorization
model. However, it is clear from this table that Class II decays are
underestimated using the perturbative value $a_2 \cong 0.2$. There is
no theoretical reason, unlike the case of Class I decays, in which
there is emission of the light meson $\pi$ \cite{6newref}, to have
approximate factorization in the case of $D(D^*)$ emission. There are
non-perturbative corrections to factorization, as pointed out in
Appendix C, that suggest an effective value for $a_2$. In Table 8 we
give the results for
$$a_1 \cong 1\ , \qquad a_2 \cong 0.3 \eqno({\rm B.8})$$

\noi that agree with the data within errors.\par \vskip 5 truemm

\begin{center}
\begin{tabular}{|c|c|c|}
\hline
Modes &Experiment &Factorization\\
$B \to D(D^*)\pi$ & &$a_1 \cong 1$, $a_2 \cong 0.3$\\
\hline
$BR \left ( \overline{B}^0 \to D^+\pi^-\right )$ &$(2.76 \pm 0.25) \times 10^{-3}$ &$(3.4 \pm 0.7) \times 10^{-3}$\\
\hline
$BR \left ( \overline{B}^0 \to D^0\pi^0\right )$ &$(2.7 \pm 0.8) \times 10^{-4}$ &$(1.3 \pm 0.9) \times 10^{-4}$\\
\hline
$BR \left ( B^- \to D^0\pi^-\right )$ &$(4.98 \pm 0.29) \times 10^{-3}$ &$(6.0 \pm 1.2) \times 10^{-3}$\\
\hline
$BR \left ( \overline{B}^0 \to D^{*+}\pi^-\right )$ &$(2.76 \pm 0.21) \times 10^{-3}$ &$(3.3 \pm 0.7) \times 10^{-3}$\\
\hline
$BR \left ( \overline{B}^0 \to D^{*0}\pi^0\right )$ &$(2.7 \pm 0.5) \times 10^{-4}$ &$(1.7 \pm 1.1) \times 10^{-4}$\\
\hline
$BR \left ( B^- \to D^{*0}\pi^-\right )$ &$(4.6 \pm 0.4) \times 10^{-3}$ &$(6.3 \pm 1.3) \times 10^{-3}$\\
\hline
\end{tabular}
\end{center}

\centerline{{\bf Table 8.} Same as Table 6 with the effective values $a_1
\cong 1$, $a_2 \cong 0.3$.}\par \vskip 5 truemm

\noi Now Class II decays are in better agreement and the overall
picture seems reasonable. In the estimation of the $B \to D^{**}\pi$
decays in Section 3 we adopt the values (B.8) for $a_1$ and $a_2$.

\section*{Appendix C. Corrections to factorization in\break\noindent  B
$\to$ D$^{\bf **}\pi$ decays.}\hspace*{\parindent}
In the simple-minded factorization approach one adopts the perturbative
Wilson coefficients $a_1 \cong 1$, $a_2 \cong 0.2$. Moreover, in this
approach, the amplitudes of Class II decays to $j = {3 \over 2}$ states
vanish in the heavy quark limit $A(\overline{B}^0 \to
D_{3/2}^{**0}\pi^0) = 0$ due to the vanishing of the decay constant
$f_{3/2}$.\par

Let us here discuss how the analysis would be modified by taking into
account non-perturbative corrections to factorization, following
Neubert \cite{3r}. In ref. \cite{3r} only the
decays $\overline{B} \to D_{3/2}^{**}\pi$ are discussed. We extend the
formalism to $\overline{B} \to D_{1/2}^{**}\pi$. \par

In the case of the $j = {1 \over 2}$ states $D^{**}$ emission is
allowed. Therefore, we write the amplitudes of Class I and Class II decays $\overline{B} \to D_{1/2}^{**}\pi$ ($J =
0,1$) in the form
$$A \left ( \overline{B}^0 \to D^{1/2\ +}_J\pi^-\right ) = a_1^{eff, 1/2} A_{fact}\left ( \overline{B}^0 \to D^{1/2\ +}_J \pi^-\right )$$
$$A \left ( \overline{B}^0 \to D^{1/2\ 0}_J \pi^0\right ) = a_2^{eff, 1/2} A_{fact}\left ( \overline{B}^0 \to D^{1/2\ +}_J \pi^-\right ) \eqno({\rm C.1})$$

\noi with
$$a_1^{eff,1/2} = \left [ c_1 (\mu) + {c_2(\mu) \over N_c}\right ] \left [ 1 + \varepsilon_1^{1/2}(\mu ) \right ] + c_2 (\mu ) \ \varepsilon_8^{1/2}(\mu )$$
$$a_2^{eff,1/2} = \left [ c_2 (\mu) + {c_1(\mu) \over N_c}\right ] \left [ 1 + \varepsilon_1^{1/2}(\mu ) \right ] + c_1 (\mu ) \ \varepsilon_8^{1/2}(\mu ) \eqno({\rm C.2})$$

\noi where the hadronic parameters $\varepsilon_1^{1/2}$,
$\varepsilon_8^{1/2}$ describing the non-factorizable contributions are
given by the matrix elements
$$<D^{1/2\ +}_J\pi^- |(\overline{d}u) (\overline{c}b)|\overline{B}^0>\ = \left [ 1 + \varepsilon_1^{1/2}(\mu) \right ] A_{fact} \left ( \overline{B}^0 \to D_{J}^{1/2 \ +}\pi^-\right )$$
$$<D^{1/2\ +}_J \pi^- |(\overline{d}t^a u) (\overline{c}t^ab)|\overline{B}^0>\ = {1 \over 2}  \ \varepsilon_8^{1/2}(\mu) \  A_{fact} \left ( \overline{B}^0 \to D_{J}^{1/2 \ +}\pi^-\right ) \ . \eqno({\rm C.3})$$

\noi We make explicit the upper script $j = {1 \over
2}$, since the situation is quite different for the decays into $j = {3
\over 2}$ states. Following \cite{1r}, we consider the large-$N_c$
counting rules
$$c_1 = 1 + O\left (1/N_c^2\right ) \qquad\qquad c_2 = O \left ( 1/N_c\right )$$
$$\ \ \varepsilon_1 = O\left ( 1/N_c^2\right ) \qquad\qquad\quad \ \  \varepsilon_8 = O\left ( 1/N_c\right ) \ .\eqno({\rm C.4})$$

\noi Keeping the terms up to order $1/N_c$ included, one finds
$$a_1^{eff,1/2}(\mu ) \cong a_1^{pert}(\mu ) \cong 1$$
$$a_2^{eff,1/2}(\mu ) \cong a_2^{pert}(\mu) + \varepsilon_8^{1/2}(\mu ) \cong 0.2 + \varepsilon_8^{1/2}(\mu) \ .\eqno({\rm C.5})$$

\noi The departures relative to the 
naive approximation presented above are given by the non-perturbative
coefficient $\varepsilon_8^{1/2} (\mu )$. These quantities
do not affect Class I decays, but only Class II and Class III [cf. the
isospin relation (1)]. \par

In the analysis of Appendix B on the well measured $B \to D(D^*)\pi$
decays, we have found $a_1^{eff} \cong 1$ and $a_2^{eff} \cong 0.3$.
Although there is no firm theoretical argument, we have adopted in the
text the same values, i.e.,  
$$a_1^{eff, 1/2} \cong 1 \qquad a_2^{eff,1/2} \cong 0.3 \ .\eqno({\rm C.6})$$

Going now to the case of $j = {3 \over 2}$ states, we define
$$A \left ( \overline{B}^0 \to D^{3/2\ +}_ J \pi^-\right ) = a_1^{eff, 3/2} \ A_{fact} \left ( \overline{B}^0 \to D^{3/2\ +}_J \pi^-\right )$$
$$- \sqrt{2}\ \left ( \overline{B}^0 \to D^{3/2\ 0}_J\pi^0\right ) = \varepsilon_8^{3/2}\ A_{fact} \left ( \overline{B}^0 \to D^{3/2\ +}_J\pi^-\right )\eqno({\rm C.7})$$

\noi where, due to (C.4),
$$a_1^{eff,3/2} = \left [ c_1 (\mu) + {c_2(\mu) \over N_c}\right ] \left [ 1 + \varepsilon_1^{3/2}(\mu ) \right ] + c_2 (\mu ) \varepsilon_8^{3/2}(\mu ) \cong a_1^{pert}(\mu ) \cong 1 \ . \eqno({\rm C.8})$$

\noi The hadronic coefficients
$\varepsilon_1^{3/2}(\mu )$, $\varepsilon_8^{3/2}(\mu )$ are defined
like in (C.3). We have kept the notation $\varepsilon_8^{3/2}$ in (C.7)
because, switching off non-perturbative corrections, this coefficient
does not have as a limit a non-vanishing perturbative coefficient,
unlike $a_1^{eff,3/2}$. Since the amplitude $A(\overline{B}^0 \to
D_{J}^{3/2\ 0}\pi^0)$ vanishes in the heavy quark limit, because $f_{3/2} = 0$, the amplitude chosen for the normalization is
the one that is allowed, the $\pi^-$ emission one.\par

Taking into account a non-vanishing coefficient $\varepsilon_8^{3/2}$
in (C.7), 
$$\Gamma \left ( \overline{B}^0 \to D_2^{3/2\ 0}\pi^0\right ) = \Gamma \left ( \overline{B}^0 \to D_1^{3/2\ 0}\pi^0\right )$$
$$= {G_F^2 \over 16 \pi} \left | V_{cb}V_{ud}^*\right |^2 {1 \over 2} \left [ \varepsilon_8^{3/2}\right ]^2 m_B^3\ f_{\pi}^2 {(1-r)^5(1+r)^7 \over 16r^3}  \left | \tau_{3/2} \left ( {1+r^2 \over 2r}\right ) \right |^2 \eqno({\rm C.9})$$

\noi where $r = {m_D^{3/2} \over m_B} \cong 0.46$. From the upper limits
(\ref{66e}) and the central value (\ref{18e}) for $\tau_{3/2} \left ( {1 + r^2 \over 2r}\right )$ we can infer an upper limit for $|\varepsilon_8^{3/2}|$, namely
$$|\varepsilon_8^{3/2}| < 0.90 \ . \eqno({\rm C.10})$$

\noi As expected, since the upper bounds (\ref{66e}) are rather loose, we
obtain a large upper bound on $|\varepsilon_8^{3/2}|$.

\section*{Appendix D. Remarks on 1/m$_{\bf Q}$
corrections.}\hspace*{\parindent}
Using the formalism of \cite{32r} and assuming factorization, one can
in principle compute the analytical expressions of the $1/m_Q$ ($Q = b$
or $c$) corrections to the rates $B \to D_J^j\pi$. Let us consider
Class I decays, $\overline{B}^0 \to D_J^{j+}\pi^-$. Many subleading
form factors contribute and, although a theoretical effort has been
made in their estimation for the $j = {3 \over 2}$ states within the
QCD Sum Rules approach \cite{33r}, we do not have presently at our
disposal an estimation of all the subleading form factors defined in
\cite{32r}. Therefore, we are not able at present to make an estimation
of these corrections. However, a {\it formal} expansion can be done for
these decays to pions, and subleading quantities can be estimated in
some approximation, as we explain now.\par

Let us consider the most important contributions at $w = w_{max} = w_0$, that
corresponds to $q^2 \cong 0$, the value for pion decays~:
$$w \cong w_0 = {m_B^2 + m_D^2 \over 2m_Bm_D} = {1 + r^2 \over 2r} \cong 1.3 \eqno({\rm D.4})$$

\noi where $m_D$ is the mass of the corresponding $D_J^j$ meson and $r = m_D/m_B$. Therefore, one can express all the mass factors in terms of $w_0$ and a common overall scale.\par

Then, for $\pi$ decays, using (D.4), there are two small parameters that
characterize the corrections to the
rates, namely
$$w_0 - 1 \cong 0.3 \qquad\qquad\hbox{and} \qquad \qquad {\Lambda_{QCD} \over 2 m_Q} \qquad (Q=c,b)\ . \eqno({\rm D.7})$$

\noi It is therefore convenient to classify the subleading corrections
to the rates as being of successive orders
$$\left ( w_0 - 1 \right )^s \left ( {\Lambda_{QCD} \over 2 m_Q}\right )^t\ . \eqno({\rm D.8})$$

\noi We
decide to retain only the subleading orders contributing to the rate
$(t=1)$ with the dominant order in $(w_0-1)$, namely $s = - 1/2$.\par

Using the formulas of ref. \cite{32r}, neglecting higher orders of the type (D.8), and keeeping the dominant order $(w_0-1)^{-1/2} \left ( {\Lambda_{QCD} \over 2m_Q}\right )$, we find
$$\Gamma \left ( \overline{B}^0 \to D_J^{+j} \pi^-\right ) \cong \Gamma_0 \left (  \overline{B}^0 \to D_J^{+j} \pi^-\right ) \left ( 1 + \delta_J^j\right )\eqno({\rm D.9})$$

\noi with
$$\delta_2^{3/2} = 0$$
$$\delta_1^{3/2} = {2\sqrt{2} \ \Delta E_{3/2} \over \sqrt{w_0-1}}\ {1 \over 2m_c}$$
$$\delta_1^{1/2} = {\sqrt{2} \ \Delta E_{1/2} \over \sqrt{w_0-1}}\ \left ( {3 \over 2m_b} - {1 \over 2m_c}\right )$$
$$\delta_0^{1/2} = {3\sqrt{2} \ \Delta E_{1/2} \over \sqrt{w_0-1}}\ \left ( {1 \over 2m_b} + {1 \over 2m_c}\right )\eqno({\rm D.10})$$

\noi and $\Gamma_0$ denotes the leading rate. Numerically, some of these terms of order $1/m_Q$ are not small, as they are respectively of the order
$$\delta_2^{3/2} = 0 \qquad \delta_1^{3/2} \cong 0.7 \qquad \delta_1^{1/2} \cong 0 \qquad \delta_0^{1/2} \cong 1.3 \eqno({\rm D.11})$$

\noi for $\Delta E_{3/2} \cong \Delta E_{1/2} \cong 0.4$~GeV, $m_c \cong 1.5$~GeV, $m_b \cong 4.8$~GeV and $w_0-1 \cong 0.3$.\par

For $j ={3 \over 2}$ states, the trend is not in the right
direction to explain the different central values (\ref{11e}). This
gives an idea of the type of uncertainties induced by the $1/m_Q$
corrections, that are large. One can guess that the extraction of
$|\tau_{1/2}(w_0)|^2$, $|\tau_{3/2}(w_0)|^2$ made in Section 2 is
uncertain by about a factor 2 due to these corrections. Therefore one
could assume a reasonable additional 40\% uncertainty on the values
(\ref{18e}) given for $|\tau_{1/2}(w_0)|$, $|\tau_{3/2}(w_0)|$.\par

Although these estimations give large corrections, this impression
could be wrong, since many subleading form factors contribute \cite{32r}, and we do not
know the magnitude or sign of the neglected terms. Moreover, we have taken only the leading order in the expansion (D.8). The aim of
this exercise has been to emphasize that the corrections in $1/m_Q$
are possibly large. If this was actually the case, this would upset the
results of the present stage of this analysis.


\begin{thebibliography}{99}
\bibitem{11r} That the $D^{**0}$ emission diagram could be very
important was pointed out to one of us (A. Le Yaouanc) by A. Bondar
from the Belle Collaboration. The argument is made explicit in ref. \cite{5r}.
\end{thebibliography}
\end{document}